\documentclass[prl,aps,twocolumn,preprintnumbers, showpacs, nofootinbib,superscriptaddress,notitlepage]{revtex4-1}
\usepackage{mathrsfs}
\usepackage{amsfonts}
\usepackage{amsmath}
\usepackage{slashed}
\usepackage{array}
\usepackage{verbatim}
\usepackage{epsfig}
\usepackage{graphicx}
\usepackage{color}
\usepackage[dvipsnames]{xcolor}
\usepackage{multirow}
\usepackage{booktabs}
\usepackage{mathrsfs}
\usepackage[normalem]{ulem}

\newcommand{\beq}{\begin{eqnarray}}
\newcommand{\eeq}{\end{eqnarray}}

\newcommand{\nn}{\nonumber}

\begin{document}

\title{
Pion and Kaon Distribution Amplitudes from Lattice QCD}

\collaboration{\bf{Lattice Parton Collaboration ($\rm {\bf LPC}$)}}

\author{Jun Hua}
\affiliation{Guangdong Provincial Key Laboratory of Nuclear Science, Institute of Quantum Matter, South China Normal University, Guangzhou 510006, China}
\affiliation{Guangdong-Hong Kong Joint Laboratory of Quantum Matter, Southern Nuclear Science Computing Center, South China Normal University, Guangzhou 510006, China}

\affiliation{INPAC, Shanghai Key Laboratory for Particle Physics and Cosmology, Key Laboratory for Particle Astrophysics and Cosmology (MOE), School of Physics and Astronomy, Shanghai Jiao Tong University, Shanghai 200240, China}

\author{Min-Huan Chu}
\affiliation{INPAC, Shanghai Key Laboratory for Particle Physics and Cosmology, Key Laboratory for Particle Astrophysics and Cosmology (MOE), School of Physics and Astronomy, Shanghai Jiao Tong University, Shanghai 200240, China}
\affiliation{Shanghai Key Laboratory for Particle Physics and Cosmology, Key Laboratory for Particle  Astrophysics and Cosmology (MOE), Tsung-Dao Lee Institute, Shanghai Jiao Tong University, Shanghai 200240, China}

\author{Jin-Chen He}
\affiliation{School of Physics and Astronomy, Shanghai Jiao Tong University, Shanghai 200240, China}
\affiliation{Department of Physics, University of Maryland, College Park, MD 20742, USA}

\author{Xiangdong Ji}
\affiliation{Department of Physics, University of Maryland, College Park, MD 20742, USA}

\author{Andreas Sch\"afer}
\affiliation{Institut f\"ur Theoretische Physik, Universit\"at Regensburg, D-93040 Regensburg, Germany}


\author{Yushan Su}
\affiliation{Department of Physics, University of Maryland, College Park, MD 20742, USA}

\author{Peng Sun} 
\affiliation{Department of Physics and Institute of Theoretical Physics, Nanjing Normal University, Nanjing, Jiangsu, 210023, China}

\author{Wei Wang}
\affiliation{INPAC, Shanghai Key Laboratory for Particle Physics and Cosmology, Key Laboratory for Particle Astrophysics and Cosmology (MOE), School of Physics and Astronomy, Shanghai Jiao Tong University, Shanghai 200240, China}

\author{Ji Xu}
\affiliation{INPAC, Shanghai Key Laboratory for Particle Physics and Cosmology, Key Laboratory for Particle Astrophysics and Cosmology (MOE), School of Physics and Astronomy, Shanghai Jiao Tong University, Shanghai 200240, China}
\affiliation{School of Physics and Microelectronics, Zhengzhou University, Zhengzhou, Henan 450001, China}

\author{Yi-Bo Yang}
\email{Corresponding author:ybyang@mail.itp.ac.cn}
\affiliation{CAS Key Laboratory of Theoretical Physics, Institute of Theoretical Physics, Chinese Academy of Sciences, Beijing 100190, China}
\affiliation{School of Fundamental Physics and Mathematical Sciences, Hangzhou Institute for Advanced Study, UCAS, Hangzhou 310024, China}
\affiliation{International Centre for Theoretical Physics Asia-Pacific, Beijing/Hangzhou, China}
\affiliation{School of Physical Sciences, University of Chinese Academy of Sciences,
Beijing 100049, China}

\author{Fei Yao}
\affiliation{Center of Advanced Quantum Studies, Department of Physics, Beijing Normal University, Beijing 100875, China}

\author{Jian-Hui Zhang}
\email{Corresponding author:zhangjianhui@bnu.edu.cn}
\affiliation{Center of Advanced Quantum Studies, Department of Physics, Beijing Normal University, Beijing 100875, China}

\author{Qi-An Zhang}
\affiliation{School of Physics, Beihang University, Beijing 102206, China}

\begin{abstract}
We present {the state-of-the-art lattice QCD calculation of the pion and kaon light-cone distribution amplitudes (DAs) using large-momentum effective theory. The calculation is done at three lattice spacings $a\approx\{0.06,0.09,0.12\}$~fm and physical pion and kaon masses, with the meson momenta $P_z = \{1.29,1.72,2.15\}$ GeV. The result is non-perturbatively renormalized in a recently proposed hybrid scheme with self renormalization, and extrapolated to the continuum as well as the infinite momentum limit.} We find {a significant deviation of the pion and kaon DAs from the asymptotic form, and a large $SU(3)$ flavor breaking effect in the kaon DA.}
\end{abstract}

\maketitle

\textit{Introduction:}\ 
Light pseudoscalar mesons play a fundamental role in quantum chromodynamics (QCD) as they are the (pseudo) Nambu-Goldstone bosons associated with dynamical chiral symmetry breaking (DCSB)~\cite{Weinberg:1996kr}, an important non-perturbative 
phenomena in the standard model (SM). 
Their internal structure and its impact on experimental measurements have been actively investigated for many years.

Among others, the leading-twist pion and kaon light-cone distribution amplitudes (DAs) are the simplest physical quantities to describe such internal structure, and provide a probability amplitude interpretation on how the longitudinal momentum of the pion/kaon is distributed among quarks in its leading Fock state. 
They are critical inputs for the description of hard exclusive reactions, such as 
the B meson weak decays~\cite{Cheng:2009cn,Su:2010vt} which provide useful information on CP violation and the Cabibbo-Kobayashi-Maskawa matrix, 
and play a crucial role for probes of new physics~\cite{LHCb:2019hip}; they are also important for the study of the pion elastic form factors~\cite{Farrar:1979aw}, the pion-photon transition form factor~\cite{Wang:2017ijn}, and of hard exclusive meson production which may give access to nucleon generalized parton distributions~\cite{Ji:1996ek,Radyushkin:1996ru}.

In the high energy limit, it is well-known that these DAs follow a simple asymptotic form $\phi(x)=6x(1-x)$~\cite{Lepage:1979zb}. However, 
their shapes at lower scales have been long debated. For example, the CZ model~\cite{Chernyak:1981zz} proposed a ``double-humped" shape for the pion DA which allowed a consistent description of the experimental data at that time, while other models~(see, e.g., \cite{RuizArriola:2006jge,Radyushkin:1994xv,RuizArriola:2002bp,Chang:2013pq,Agaev:2012tm}) do not favor such a structure. In addition, the 
measured electromagnetic form factors of pion/kaon at relatively large momentum transfer also exhibit some puzzling feature that might be connected to their non-asymptotic behavior~\cite{Shi:2014uwa, Holt:2012gg}. Theoretical calculations from lattice QCD will be able to shed more light on the shape of these DAs. 

There have been a lot of lattice studies on the pion/kaon DA using the traditional moments approach~\cite{Gockeler:2005jz,Braun:2006dg,Boyle:2006pw,Arthur:2010xf,Braun:2015axa,Bali:2017ude,RQCD:2019osh}. The recently proposed large-momentum effective theory (LaMET)~\cite{Ji:2013dva,Ji:2014gla,Ji:2020ect} allows to access the entire $x$-dependence of the DAs from first principle lattice calculations, instead of only the first few moments (for other proposals with applications to the DAs, see Refs.~\cite{Braun:2007wv,Bali:2017gfr,Bali:2018spj,Detmold:2021qln}). 
Using LaMET, several calculations of the $x$-dependence of meson DAs have been carried out~\cite{Zhang:2017bzy,Zhang:2017zfe,Zhang:2020gaj}. However, a recent analysis~\cite{Zhang:2020rsx} showed that the nonperturbative renormalization of the quasi-light-front (quasi-LF) correlation in LaMET could be highly non-trivial, especially when off-shell quark matrix elements are used. In such a case, even after renormalization there may still be residual linear divergences rendering the continuum extrapolation problematic. 
To resolve this issue, a self renormalization strategy~\cite{LPC:2021xdx} has been proposed, where one fits the divergence structure to a quasi-LF correlation and uses it for renormalization.
The present work provides the first full implementation of this strategy, and shows that it indeed gives promising results.

\textit{Lattice simulation:}\
%
The leading-twist light-cone DA of a pseudoscalar meson is defined as 
\begin{eqnarray} 
&& \int \frac{d\xi^-}{2\pi} e^{ixp^+ \xi^-}\langle 0 | \bar \psi_1(0) n\!\!\!\slash \gamma_5 U(0,\xi^-)\psi_2(\xi^-) |M(P)\rangle  \nonumber \\
&&=i f_M(p \cdot n) \phi_{M}(x), 
\end{eqnarray} 
where $U(0,\xi^-) = P\,{\rm exp}\big[ig_s\int_{\xi_-}^0 ds ~ {n}\cdot A(sn)\big]$ is the path-ordered gauge link defined along the minus light-cone direction $n$ ($n^2=0$).
To extract this quantity, 
we calculate the following quasi-LF correlation on the lattice with momentum $\vec{P}$ along the $z$-direction~\cite{Zhang:2017bzy}: 
\begin{align}
 C_{2}^{m}(z,\vec P, t)  &= \int d^3y e^{-i\vec P\cdot \vec y} \nonumber \\ \nonumber
  & \times \langle 0 |{\cal O}_{\Gamma_1}(z;\vec{y},t) \bar  \psi_2(0,0) \Gamma_2 \psi_1(0,0)|0\rangle, 
 \label{eq:twopt}
\end{align} 
where ${\cal O}_{\Gamma_1}(z;\vec{y},t)\equiv\bar \psi_1(\vec y, t) \Gamma_1 U(\vec y, \vec y {\color{blue}-} z\hat z) \psi_2(\vec y {\color{blue}-} z\hat z, t)$ is the quasi-LF operator with $\hat z$ being the unit vector in the $z$-direction, $U(\vec{x}, \vec{x} {\color{blue}-} \vec z)$ is the spatial Wilson line connecting lattice sites $\vec{x}$ and $\vec{x} - \vec z$, $\psi_2\Gamma_2 \psi_1$ is the interpolating field of the meson $m$, and $\Gamma_{1,2}$ are chosen as $\Gamma_1 = \gamma_z \gamma_5,\, \Gamma_2=\gamma_5$ for the pseudoscalar meson. The ground-state matrix elements can be extracted from the following two-state fit formula:
\begin{eqnarray}
\frac{ C^{m}_{2}(z,\vec P, t)  }{ C^{m}_{2}(z=0,\vec P, t)  } = \frac{ H_{m}^{\rm B}(z)(1+ c_{m}(z) e^{-\Delta Et}) }{(1+ c_{m}(0) e^{-\Delta Et}) }, \label{eq:c2_ratio}
\end{eqnarray}
where $H_{m}^{\rm B}(z)$ is the normalized ground-state matrix element, $c_m$ and $\Delta E$ are free parameters accounting for (one or more) excited state contamination, which are exponentially suppressed in the large time limit. 
Based on the comparison of one- and two-state fits (see supplemental material~\cite{supplemental}), we use the one-state fit results in the analysis below with $t_{min}=0.72,0.54,0.42$ fm (for $P_z=1.29,1.72,2.15$~GeV) which is large enough to eliminate the excited states contamination.

\begin{table}
\centering
\caption{Details of the simulation setup. The light and strange quark mass (both valence and sea quark) of the clover action are tuned such that $m_{\pi}$=140 MeV and $m_{\eta_s}$=670 MeV.}
\label{Tab:setup}
\begin{tabular}{cclccccccc}
\hline
\hline
Ensemble ~~& $a$(fm) & \ \!$L^3\times\  T$   & $c_{\mathrm{SW}}$ & $m_{u/d}$  & $m_{s}$  \\
\hline
a06m130  ~~& 0.057  ~~& $96^3\times$192   ~~& 1.03493  ~~&-0.0439         ~~&-0.0191            \\ 
a09m130  ~~& 0.088  ~~& $64^3\times$~ \!96  ~~& 1.04239  ~~&-0.0580         ~~&-0.0174           \\ 
a12m130  ~~& 0.121  ~~& $48^3\times$~ \!64  ~~& 1.05088  ~~&-0.0785         ~~&-0.0191           \\
\hline
\end{tabular}
\end{table}
In this work, the simulation is done using the clover fermion action on three ensembles with 2+1+1 flavors of highly improved staggered quarks  (HISQ) generated by the MILC collaboration \cite{Follana:2006rc, MILC:2012znn}, at physical pion mass with three lattice spacings $0.057, 0.088$ and $0.121$~fm.  Hypercubic (HYP) smeared fat links~\cite{Hasenfratz:2001hp} are used in both the fermion action and the quasi-LF operators in $C_2^m$ to improve the signal-to-noise ratio. The rest of the simulation setup is summarized in Table~\ref{Tab:setup}. In addition, we use momentum smeared 2-2-2 grid sources, repeat the calculation at several time slices, and average the forward and backward correlation functions to improve statistics. In total, we have $570\ ({\textrm{cfg.}})\times 8\ (\textrm{grid source}) \times 8\ (\textrm{source time slices}) \times 2\ (\textrm{forward/backward})$, $730\times 8 \times 6 \times2$ and $970 \times 8 \times 4 \times2$ measurements at three ensembles with $a=0.057$, 0.088 and 0.121 fm, respectively.

\textit{Hybrid scheme with self renormalization:}
The bare quasi-LF correlation calculated above contains both linear and logarithmic ultraviolet (UV) divergences which have to be removed by renormalization. On the lattice, the numerical subtraction of linear divergences is extremely delicate. In particular, such divergences may not be fully removed if the RI/MOM renormalization scheme is used~\cite{Zhang:2020rsx}. Here we adopt the self renormalization proposed in Ref.~\cite{LPC:2021xdx}, which amounts to fitting the bare quasi-LF correlation and subtracting the relevant UV divergences. To be more precise, one fits the bare quasi-LF correlation at given hadron momentum and multiple lattice spacings with a perturbative-QCD-dictated parametrization that contains a linear divergence, a logarithmic divergence, and discretization effects. After removing all the UV divergences and discretization effects, one is left with the renormalized quasi-LF correlation encoding the intrinsic non-perturbative physics. 

As suggested in Ref.~\cite{LPC:2021xdx}, the UV divergences in the quasi-LF correlator can be determined by using, e.g., the pion PDF matrix elements $\mathcal{M}(z)\equiv \langle \pi|{\cal O}_{\gamma_t}|\pi\rangle $ in the rest frame at multiple lattice spacings, 
and fitting the {bare} data \textbf{$\mathcal{M}^{\rm B}$} to the following form~\cite{LPC:2021xdx} 
{
\begin{eqnarray}
&\mathcal{M}^{\rm B}(z,a) = Z^{\rm self}(z,a) \mathcal{M}^{\rm R}(z),
\end{eqnarray}
{with the renormalization factor parametrized as}~\cite{LPC:2021xdx} 
\begin{eqnarray}
&Z^{\rm self}(z,a)\equiv\mathrm{exp}\Big\{ \frac{k z}{a \ln[a \Lambda_{\rm QCD}]} 
+ m_0 z + f(z) a\nonumber\\
&+ \frac{3 C_{F}}{b_0} \ln \Big[\frac{\ln [1 /(a \Lambda_{\rm QCD})]}{\ln [\mu / \Lambda_{\rm QCD}]}\Big]+\ln\Big[1+\frac{d}{\ln (a \Lambda_{\rm QCD})})\Big]\Big\}, \label{eq:lnM}
\end{eqnarray}
where the first term in the curly bracket is the linear divergence, $m_{0}$ denotes a finite mass contribution arising from renormalon ambiguity, etc., and $f(z) a$ accounts for the discretization effects (The $\mathcal{O}(a)$ correction here arises from the mixed action effect in using clover valence fermions on HISQ sea ones).} The last two terms come from the resummation of leading and sub-leading logarithmic divergences, which only affect the overall normalization at different lattice spacings. {To partially account for higher-order perturbative effects as well as remaining lattice artifacts, we also treat $d$ and $\Lambda_{\rm QCD}$ as fitting parameters}~\cite{LPC:2021xdx}. The renormalized matrix element is required to be equal to the continuum perturbative $\overline{\rm MS}$ result at short distances (chosen to be $z\in z_s=[0.06, 0.18]$\,fm as defined in \cite{Ji:2020brr}),
\begin{align}
&\mathcal{M}^{\rm R}(z)|_{z \in z_s}=\mathcal{M}^{\overline{\rm MS} ,\ {\rm 1-loop}  }(z)\nn\\
&\equiv 1+\frac{\alpha^{\overline{\rm MS}}_s C_F}{4\pi}\left[3 \ln\frac{z^2\mu^2}{4e^{-2\gamma_E}} + 5\right]+{\mathcal O}(\alpha^{2,\overline{\rm MS}}_s), 
\end{align}
which helps the determination of $m_0$ and $d$. In the calculation we use the $\overline{\rm MS}$ renormalization scale ${\mu=2\ \mathrm{GeV}}$ and $\Lambda_{\overline{\rm MS}}=0.24\ \mathrm{GeV}$.


In the present case, we follow the same strategy as above, except that 
the renormalized matrix element in the $\overline{\rm MS}$ scheme
\begin{equation}
H^{\overline{\rm MS}}_m(z)=H^{\rm B}_m(z,a)/ \tilde{Z}^{\rm self}(z,a),
\end{equation} 
is now required to be matched to the continuum perturbative $\overline{\rm MS}$ result of the normalized quasi-DA matrix element at short distances in the rest frame, which reads at one-loop
\beq
H^{\overline{\rm MS},\ {\rm 1-loop} }_m(z)\equiv 1 + \frac{\alpha^{\overline{\rm MS}}_s C_F}{4\pi} \left[3 \ln\frac{z^2\mu^2}{4e^{-2\gamma_E}} + 7\right].
\eeq
$\tilde{Z}^{\rm self}$ turns out to be the same as $Z^{\rm self}$ except for the value of $d$.

\begin{figure}[!th]
\begin{center}
\includegraphics[width=0.47\textwidth]{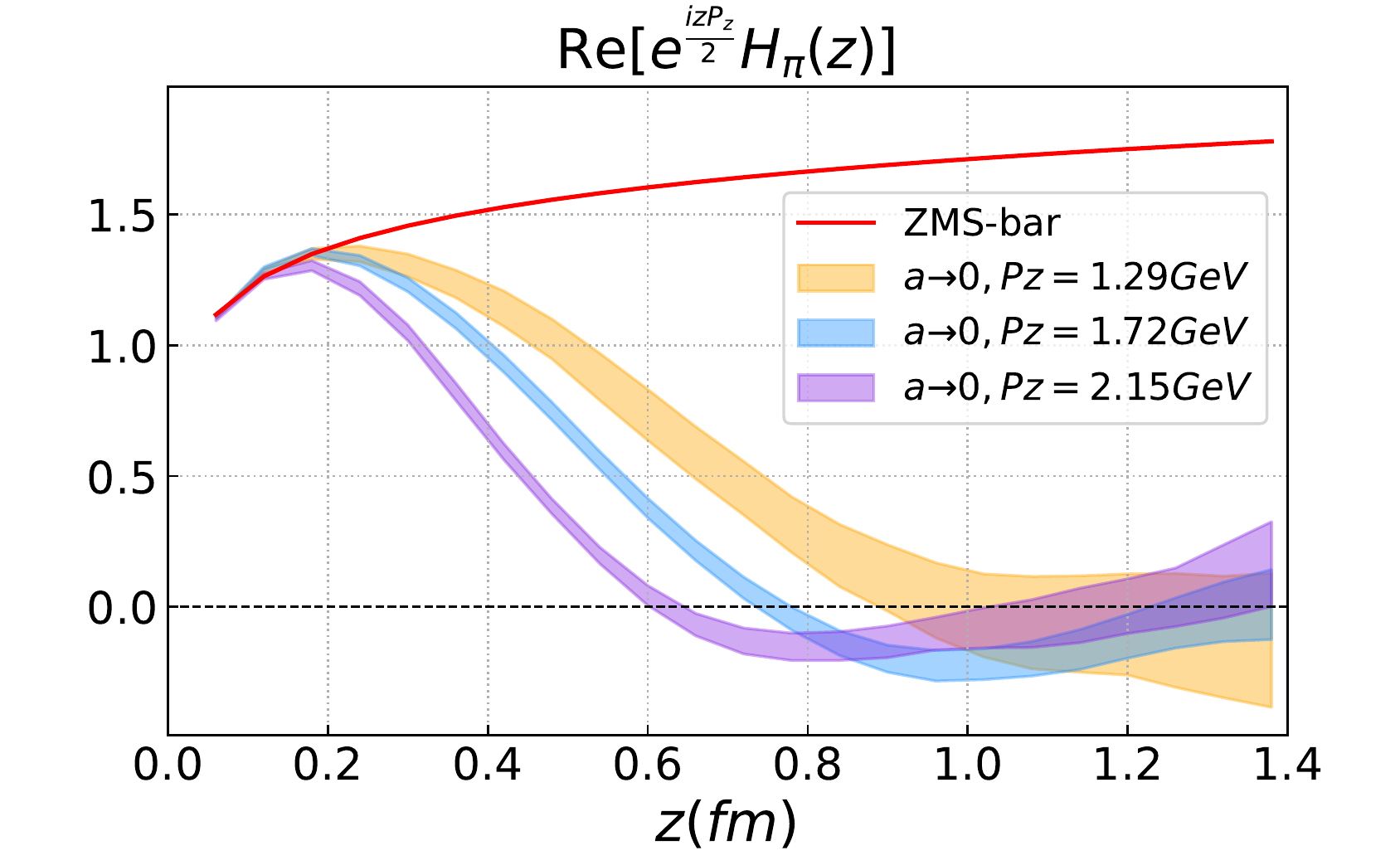}
\caption{Comparison of self-renormalized quasi-LF correlationRe[$e^{\frac{i z P_{z}}{2}}H^{\overline{\rm MS}}_m(z)$] of the pion with different momenta (bands), and the perturbative result in the $\overline{\rm MS}$ scheme $H^{\overline{\rm MS},\ {\rm 1-loop}}_m(z)$ (red curve).}\label{fig:comp_MSbar}
\end{center}
\end{figure}

In Fig.~\ref{fig:comp_MSbar}, we show a comparison between the self renormalized quasi-LF correlations Re[$e^{\frac{izP_{z}}{2}} H^{\overline{\rm MS}}_m(z)$] (after linear ${\cal O}(a)$ continuum extrapolation and phase rotation) 
with the perturbative one-loop result $H^{\overline{\rm MS},\ {\rm 1-loop}}_m(z)$. As can be seen from the figure, all quasi-LF correlations agree well with the perturbative result for small $z$, indicating a mild $P_z$ dependence in that region.  

It is worth pointing out that the self renormalization strategy above does not apply at very small $z$ due to finite lattice
spacing artifacts in the data. In the ratio scheme~\cite{Orginos:2017kos}, 
some degree of cancellation happens in the bare correlations between large momentum states and non-perturbative
lattice renormalization factors. However, in the present case, the agreement of the self-renormalized LF-correlation with the perturbative result extends down to $z\sim 0.06\,\rm{fm}$ which is our smallest lattice spacing. Thus, we only need to supplement it with the renormalized quasi-LF correlation at $z=0$ which is normalized to $1$. In this way, we obtain the fully renormalized quasi-LF correlation. To facilitate the subsequent matching procedure, we define a modified renormalized correlation by further dividing out the perturbative factor $H^{\overline{\rm MS},\ {\rm 1-loop}}_m(z)$ so that the ratio scheme matching applies,
\begin{align}
    H^{\rm R}_m(z) = \frac{H^{\rm B}_m(z,a)}{\tilde{Z}^{\rm self}(z,a) \cdot H^{\overline{\rm MS},\ {\rm 1-loop}}_m(z)}=\frac{H^{\overline{\rm MS}}_m(z)}{H^{\overline{\rm MS},\ {\rm 1-loop}}_m(z)}.
\end{align}
Note that this is equivalent to using the hybrid renormalized quasi-LF correlation and matching, as the perturbative difference in the quasi-LF correlation is exactly compensated by that in the matching.
    
From Fig.~\ref{fig:comp_MSbar}, we can see that the uncertainty of the renormalized quasi-LF correlation grows rapidly at large distance. A brute-force truncation of the correlation introduces unphysical oscillations \cite{Zhang:2020gaj} in momentum space after Fourier transformation. To resolve this issue, we adopt a physics-based extrapolation form~\cite{Ji:2020brr} at large quasi-LF distance ($\lambda=z P_z$):
\begin{align}
	  H^{\rm R}_{m}(z, P_z) &= \Big[\frac{c_1}{(i\lambda)^a} + e^{-i\lambda}\frac{c_2}{(-i \lambda)^b}\Big]e^{-\lambda/\lambda_0},
	  \label{eq:extra}
\end{align} 
where the algebraic terms in the square bracket account for a power law behavior of the DAs in the endpoint region and the exponential term comes from the expectation that at finite momentum ($\vec{P}$) the correlation function has a finite correlation length (denoted as $\lambda_0$), which becomes infinite when the momentum goes to infinity. In this work, the Lorentz boost factor $\gamma$ for the pion at the physical point is very large $\{9.21, 12.29, 15.36\}$, and thus the correlation length is very large. We therefore drop the $e^{-\lambda/\lambda_0}$ factor, and directly perform a polynomial extrapolation as suggested in~\cite{Ji:2020brr}. The details of this extrapolation can be found in the supplemental material~\cite{supplemental}.

\begin{figure}[!th]
\begin{center}
\includegraphics[width=0.47\textwidth]{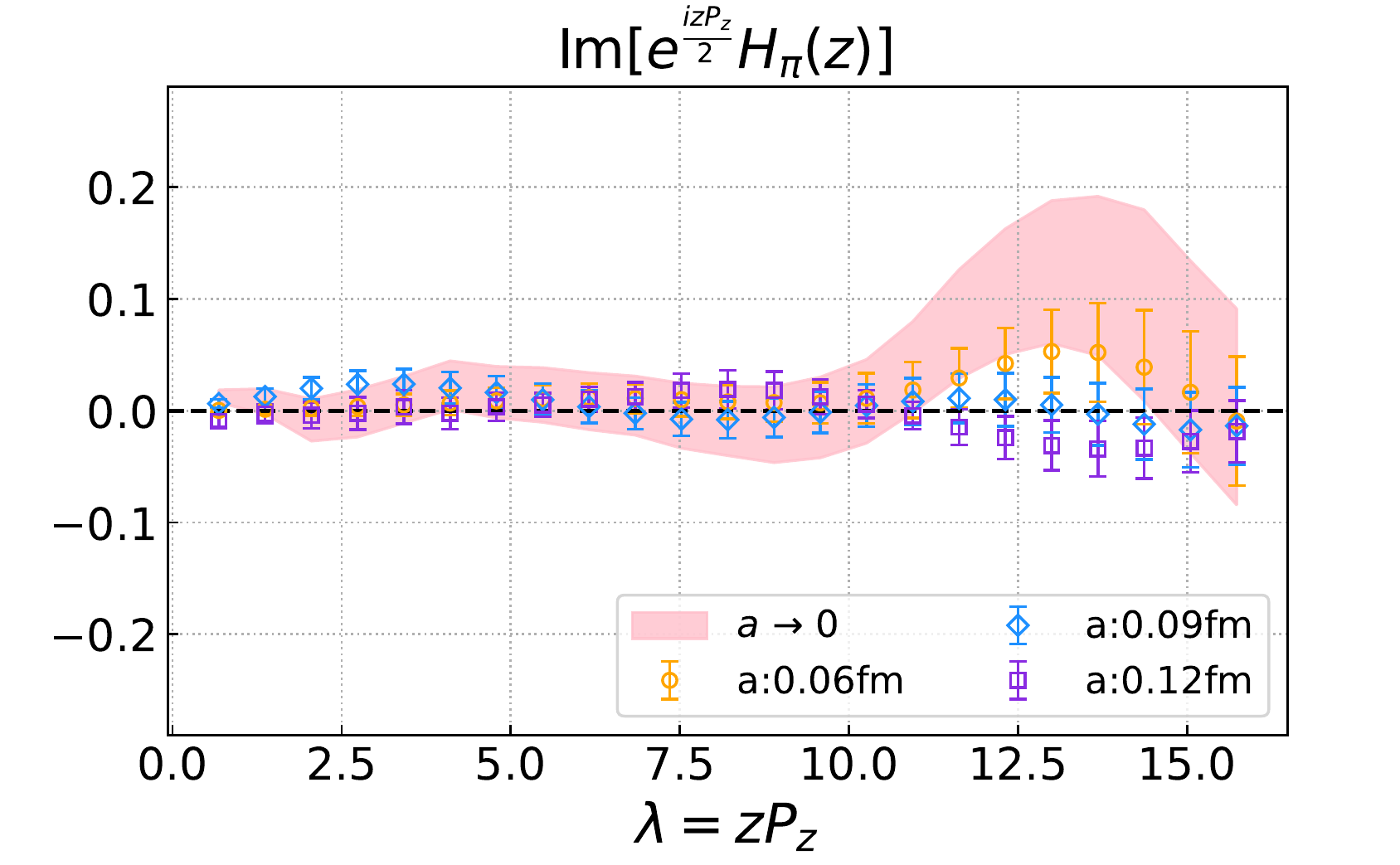}
\includegraphics[width=0.47\textwidth]{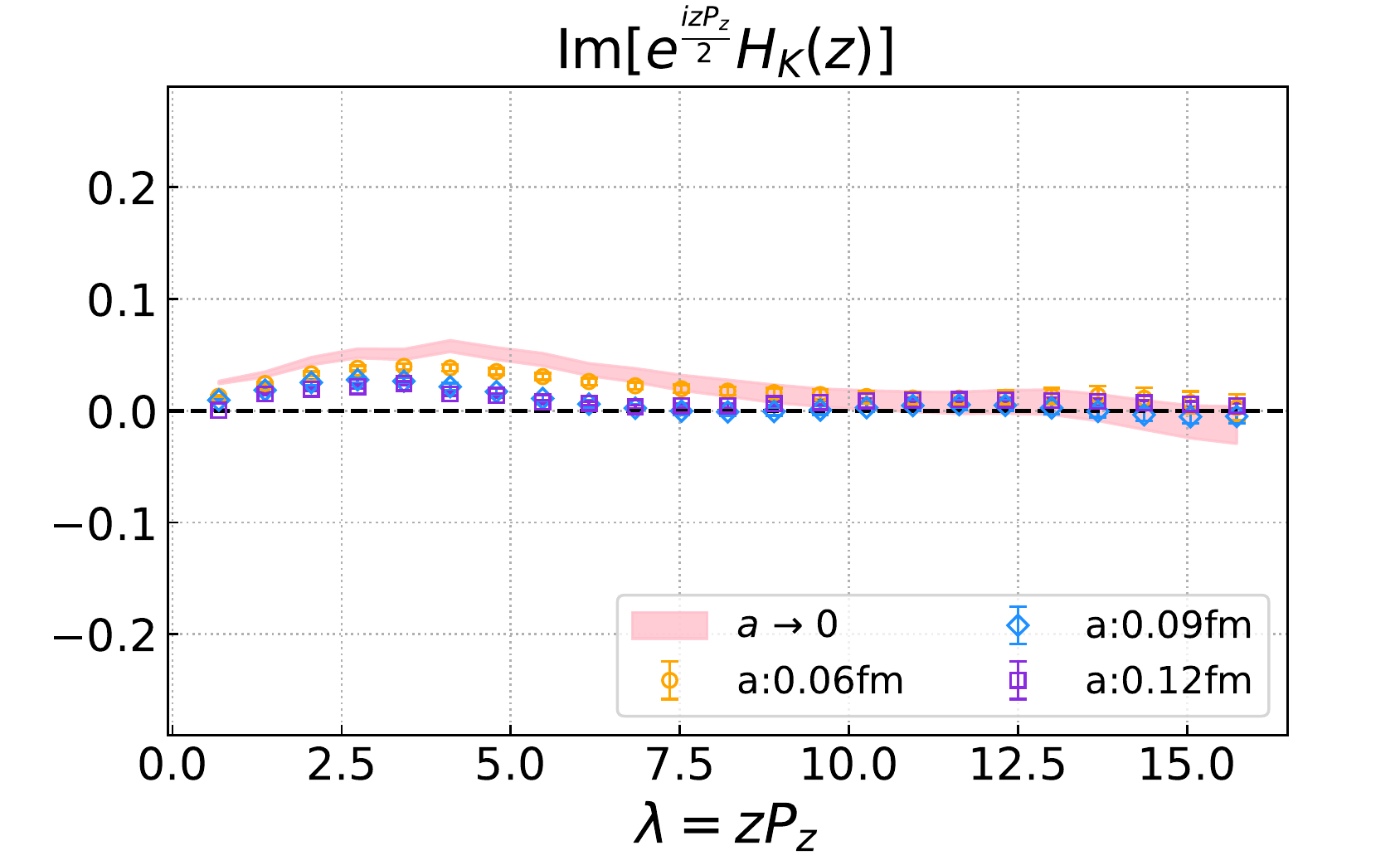}
\caption{The imaginary part of the quasi-LF correlation function ($e^{izP_z/2}H^{\rm R}_m(z)$) for the pion (top) and kaon (bottom) in the continuum limit $a \to 0$. The hadron momentum is $P_z=2.15$~GeV. }\label{fig:rota_matrix}
\end{center}
\end{figure}

\textit{Numerical results:}
In Fig.~\ref{fig:rota_matrix} we show, as an example, the imaginary part of $e^{izP_z/2}H^{\rm R}_m(z)$ for the pion (upper panel) and kaon (lower panel) at different lattice spacings with $P_z=2.15$ GeV. It reflects the SU(3) flavor breaking effects between the valence quarks in the light meson. For the pion it is consistent with zero within errors as expected, since we have used degenerate valence $u/d$ quark masses in the ensembles. While in the case of kaon there is a non-vanishing imaginary part. Such an imaginary part increases slightly with $P_z$, as observed in previous DA studies using LaMET \cite{Ji:2020brr,Zhang:2020gaj}, and a comparison of the results at different momenta can be found in the supplemental material~\cite{supplemental}.

The factorization can be done either in momentum space~\cite{Ji:2015qla,Liu:2018tox} or in coordinate space. 
Here we choose the latter, which results in
\begin{align}\label{eq:factorization}
H_m^{\rm R}(z,\lambda,\mu_R)&=\int_0^1 dx dy\,\theta(1-x-y)\nonumber\\
&\times C(x,y,z^2,\mu_R,\mu)h_m^{\rm R}(x,y,\lambda,\mu)\nonumber\\
&+ \mathcal{O} \left(\Lambda^2_{QCD}z^2,M^2 z^2\right),
\end{align}
where we take renormalization scale and factorization scale to be the same and set $\mu$ = $\mu_{R}$ = 2 GeV in this paper. $h_m^R$ is the LF correlation related to the light-cone DA through the following Fourier transformation
\begin{equation}
h_m^{\rm R}(x,y,\lambda,\mu)=\int_0^1 du\, e^{i u(x-1) \lambda-i(1-u)y\lambda}\phi(u,\mu).
\end{equation}
\begin{figure}[htbp]
\begin{center}
\includegraphics[width=0.47\textwidth]{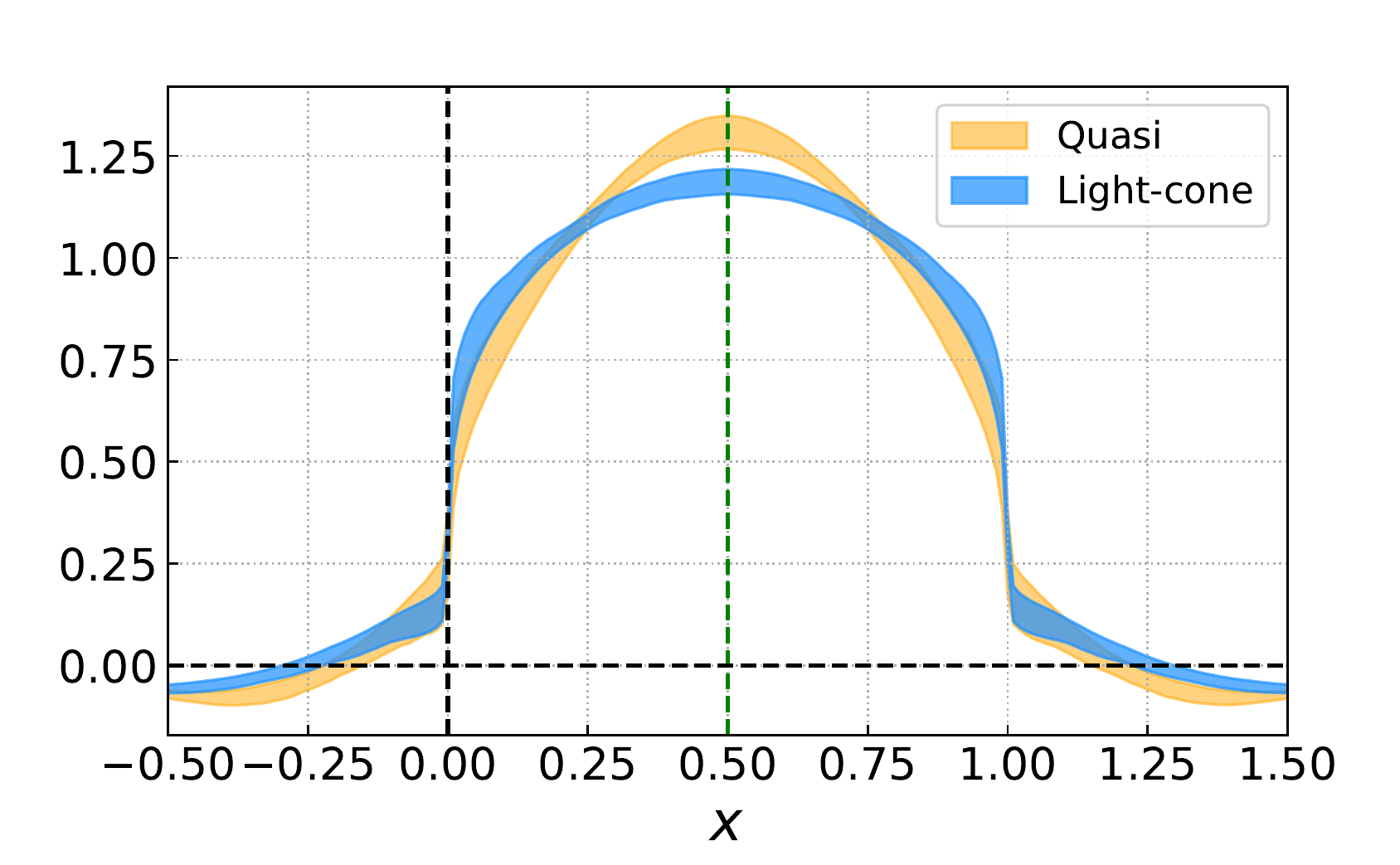}
\caption{Quasi-DA and LCDA for the pion in momentum space in the continuum limit $a \to 0$, $P_z=2.15$ GeV. }\label{fig:matching}
\end{center}
\end{figure}

\noindent The perturbative matching kernel $C$ up to the next-to-leading order is given in the supplemental material~\cite{supplemental}.

The impact of the perturbative matching is illustrated in Fig.~\ref{fig:matching}, where a Fourier transformation to momentum space has been performed. As can be seen from the figure, the matching broadens the quasi-DA in the physical region. Outside the physical region ($x<0$ or $x>1$), there still exists a non-vanishing tail, indicating potential effects of higher-order matching and higher-twist contributions. 
Nevertheless, in the unphysical region, the results are consistent with zero within $\sim 2$ standard deviations.

\begin{figure}[!th]
\begin{center}
\includegraphics[width=0.43\textwidth]{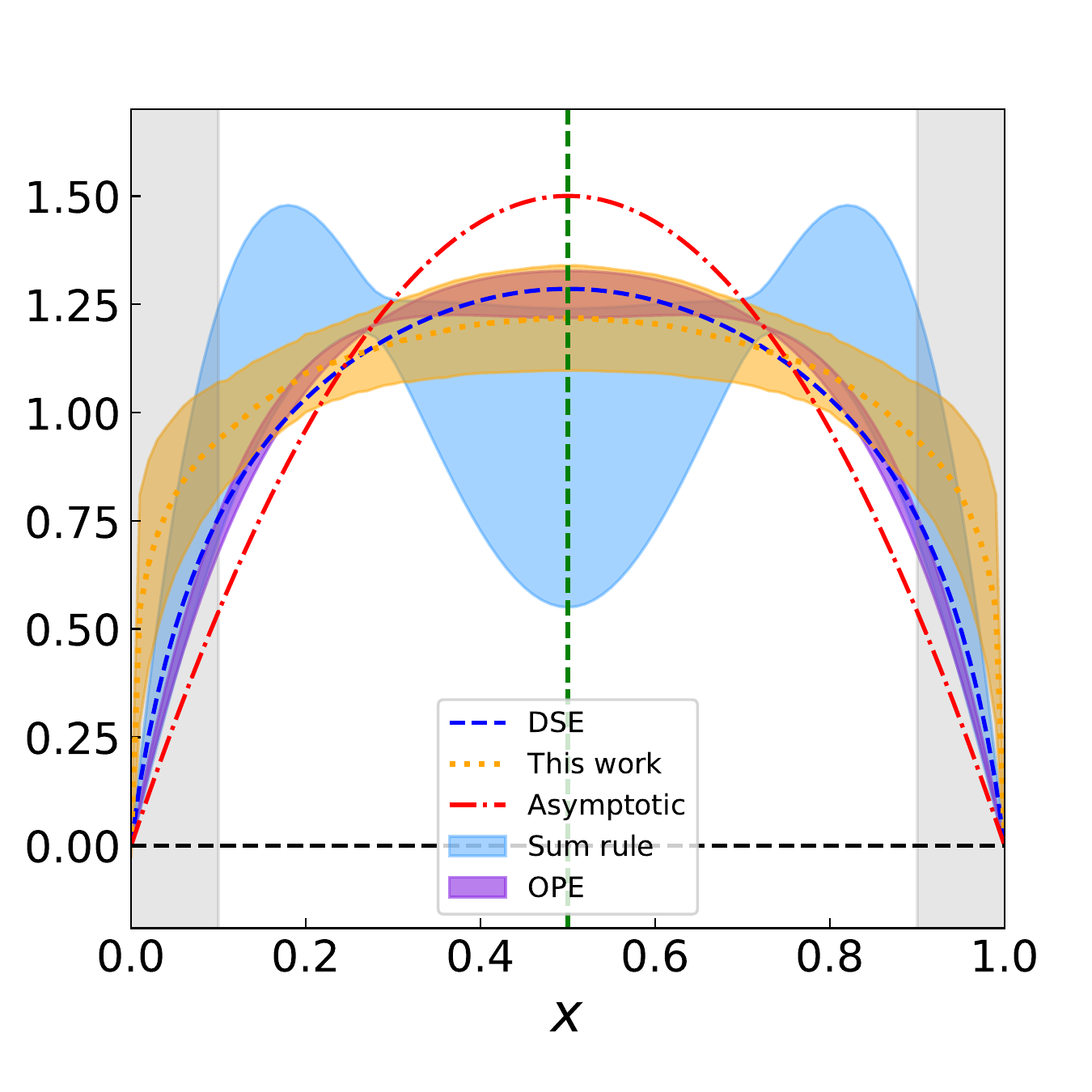}
\includegraphics[width=0.43\textwidth]{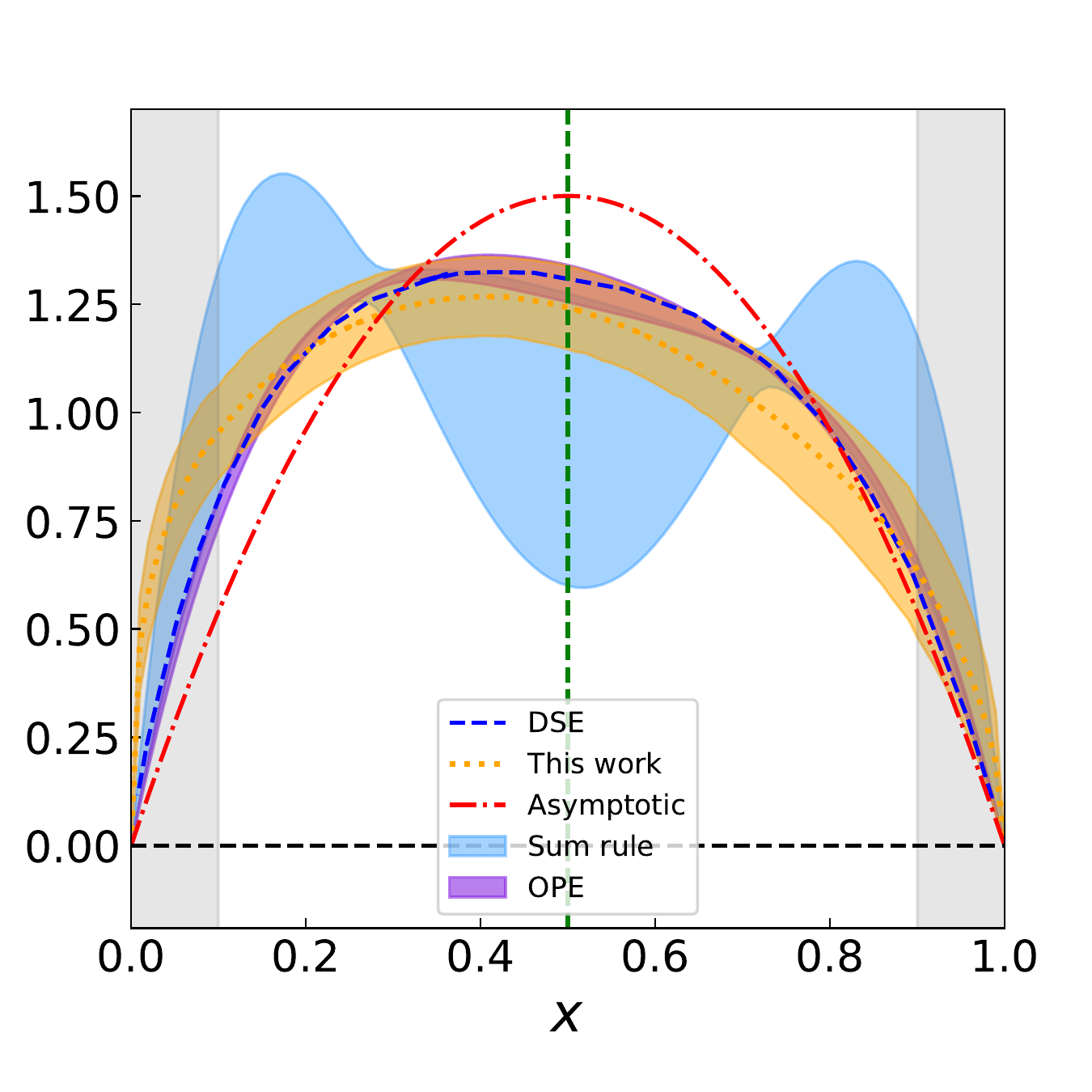}
\caption{LCDAs for $\pi$(top) and $K$(bottom), extrapolated to the continuum $(a\to 0)$ and infinite momentum limit $(P_z \to \infty)$. For the kaon, $x$ is the momentum fraction carried by the light quark.}\label{fig:lcdapik}
\end{center}
\end{figure}

With the results for $P_z = 1.29, 1.72, 2.15$ GeV above, we can perform an extrapolation to $P_z \to \infty$ using the functional form:
\begin{eqnarray}
	\phi(x, P_z) = \phi(x, P_z \to \infty)  + \frac{c_2(x)}{P^2_z} +   \mathcal{O}\Big(\frac{1}{P^4_z}\Big).
\end{eqnarray}
The final results of the $\pi, K$ DAs are given in Fig.~\ref{fig:lcdapik}, where systematic uncertainties from renormalization scale, algebraic extrapolation, continuum and infinite momentum extrapolation have been taken into account. 
As the endpoint regions can not be reliably predicted by LaMET, we adopt a phenomenological $x^a(1-x)^b$ extrapolation in this region (taken as $0<x<0.1 \,\, \& \,\, 0.9<x<1$). 
For comparison, we also plot the asymptotic form $6x(1-x)$ and results from QCD sum rules~\cite{Ball:2007zt}, Dyson-Schwinger equations (DSE)~\cite{Roberts:2021nhw}, and reconstructed from moments calculations (OPE)~\cite{RQCD:2019osh}. As can be seen from the figure, both $\pi$ and $K$ DAs deviate significantly from the asymptotic form, but are close to the results from DSE and OPE calculations. The shape of $\pi$ DA is much broader than the asymptotic form which could be explained as a direct expression of DCSB, as discussed in~\cite{Chang:2013pq}. The relatively large uncertainties in Fig.~\ref{fig:lcdapik} are expected to be reduced if the analysis of higher-order and higher-twist effects becomes available in the future.

\textit{Summary:}
We present a state-of-the-art lattice calculation of $\pi$ and $K$ DAs using LaMET. The renormalization is done in the hybrid scheme with self renormalization proposed recently. Based on the results at physical light and strange quark masses with three lattice spacings and momenta, we perform an extrapolation to the continuum and infinite momentum limit. The final results exhibit a significant deviation from the asymptotic form, while they are close to the DSE and OPE results, especially in the middle $x$ region where our method is reliable. However, there are still some significant differences in the endpoint regions. This could be due to missing higher-power/ high-order corrections in LaMET which can be improved in future calculations, or due to effects of higher moments ignored in the OPE and DSE calculations.

\section*{Acknowledgement}
The calculations were performed using the Chroma software suite~\cite{Edwards:2004sx} with QUDA~\cite{Clark:2009wm,Babich:2011np,Clark:2016rdz} through HIP programming model~\cite{Bi:2020wpt}. This work is supported in part by the Strategic Priority Research Program of Chinese Academy of Sciences (No. XDB34030302 and XDPB15), the National Natural Science Foundation of China (NNSFC) under Grants No.11735010, 11975051, 12005130, 012047503, a NSFC-DFG joint grant under grant No. 12061131006 and SCHA~458/22, and Natural Science Foundation of Shanghai under grant No. 15DZ2272100. XJ is supported partially by
the US DOE, Office of Science, grant DE-SC0020682. The computations were performed on the CAS Xiandao-1 computing environment, and also HPC Cluster of ITP-CAS. 

\clearpage

\appendix
\section{Additional information on the data analyses}

\textit{Dispersion relation:} 
The effective mass can be extracted by the two-point correlation function at z=0 through a two-state fit. The dispersion relation of $\pi$ and $K$ can be derived from the parametrized form :
\begin{eqnarray}
E^2 = m^2 + c_2(P_z)^2+c_3(P_z)^4a^2.
\end{eqnarray}
Here $C_{3,\pi} = -0.178 (19)$, $C_{3,K} =  -0.151 (13)$ parameterize the discretization effects. $C_{2,\pi} = 0.9921 (61)$, $C_{2,K} =  1.0210 (62)$ are consistent with the speed of light within $3\sigma$.

\textit{Excited state contamination:} The common approach to extract the ground-state matrix elements $H^{\rm B}_{m}(z)$ is the two-state fit in Eq.(2) of the main text. However, to correct for excited state contaminations in this way requires high precision. An alternative solution is to use a one-state fit. As the excited state contaminations get suppressed when $t$ becomes large, we can eliminate them by using a large enough $t$ range.

\begin{figure}[!th]
\begin{center}
\includegraphics[width=0.47\textwidth]{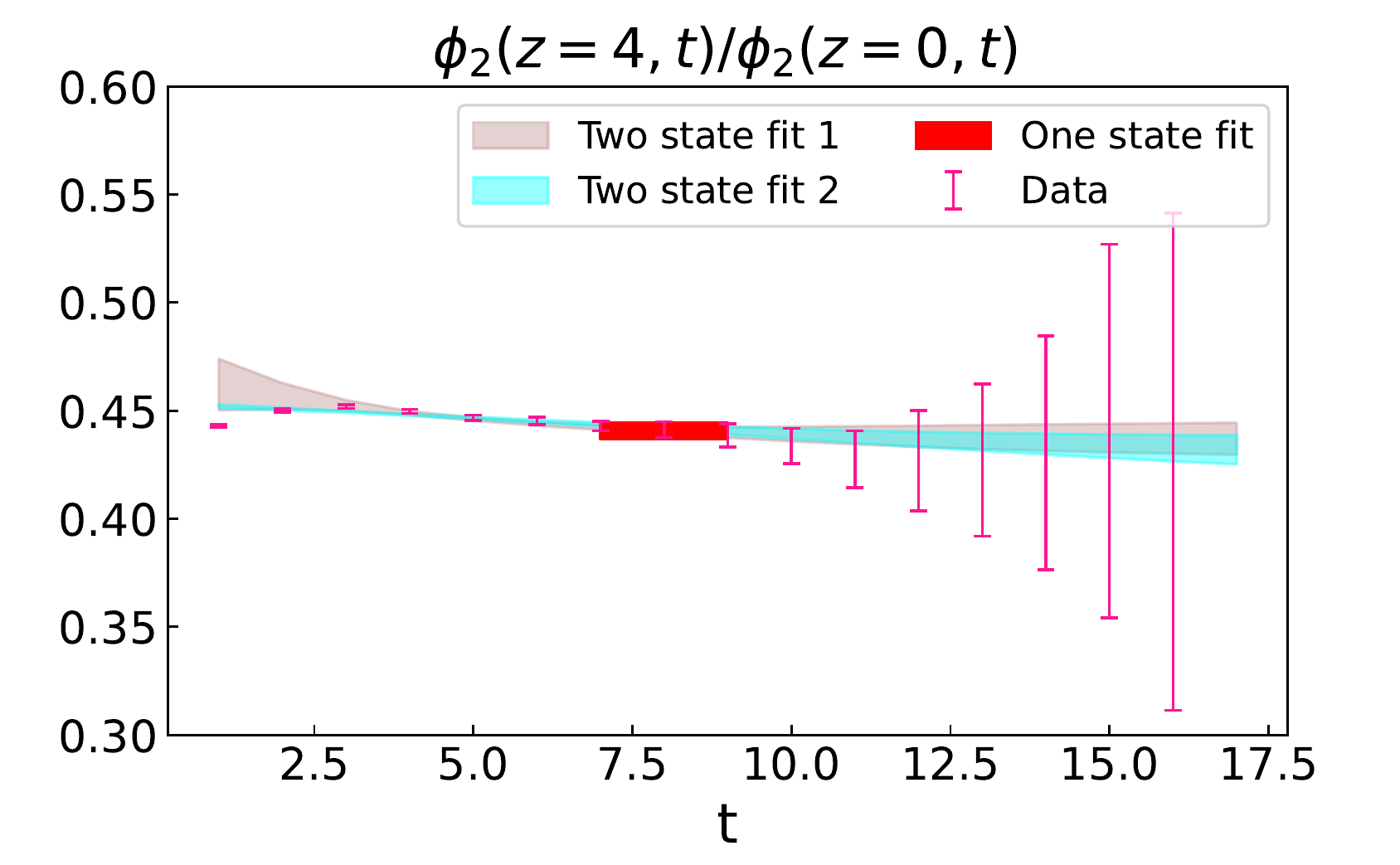}
\caption{ Comparison of two-state fit (fit 1 $\&$ fit 2) and one-state fit.  Fit 1 has  free parameters while fit 2 is constrained by our dispersion relation.}\label{fig:fitcompare}
\end{center}
\end{figure}

As shown in  Fig.~\ref{fig:fitcompare}, the excited state introduces a slight bend to the two-state fitted curve, especially in the small $t$ region. However, with current accuracy, the excited state contaminations cannot be fitted effectively, and the one-state fit result is  consistent with the two-state fit ones within statistical uncertainties.

\begin{figure}[!th]
\begin{center}
\includegraphics[width=0.47\textwidth]{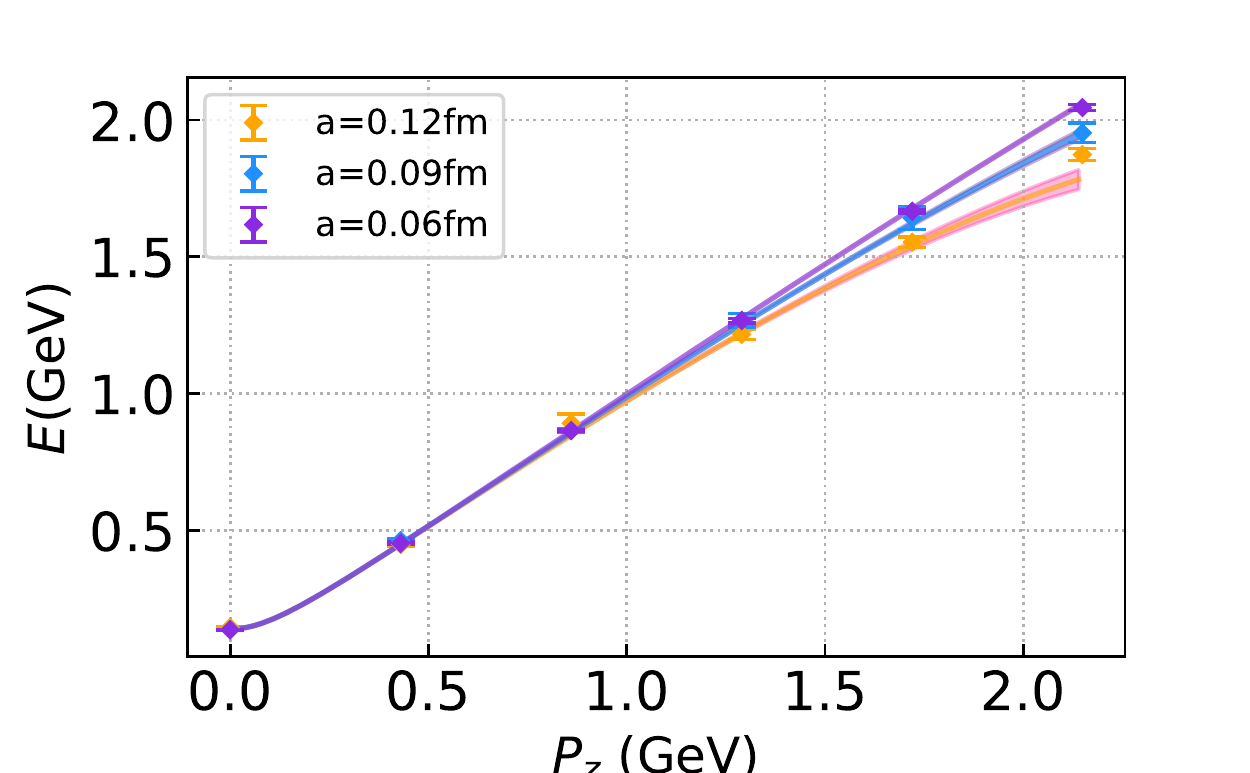}
\includegraphics[width=0.47\textwidth]{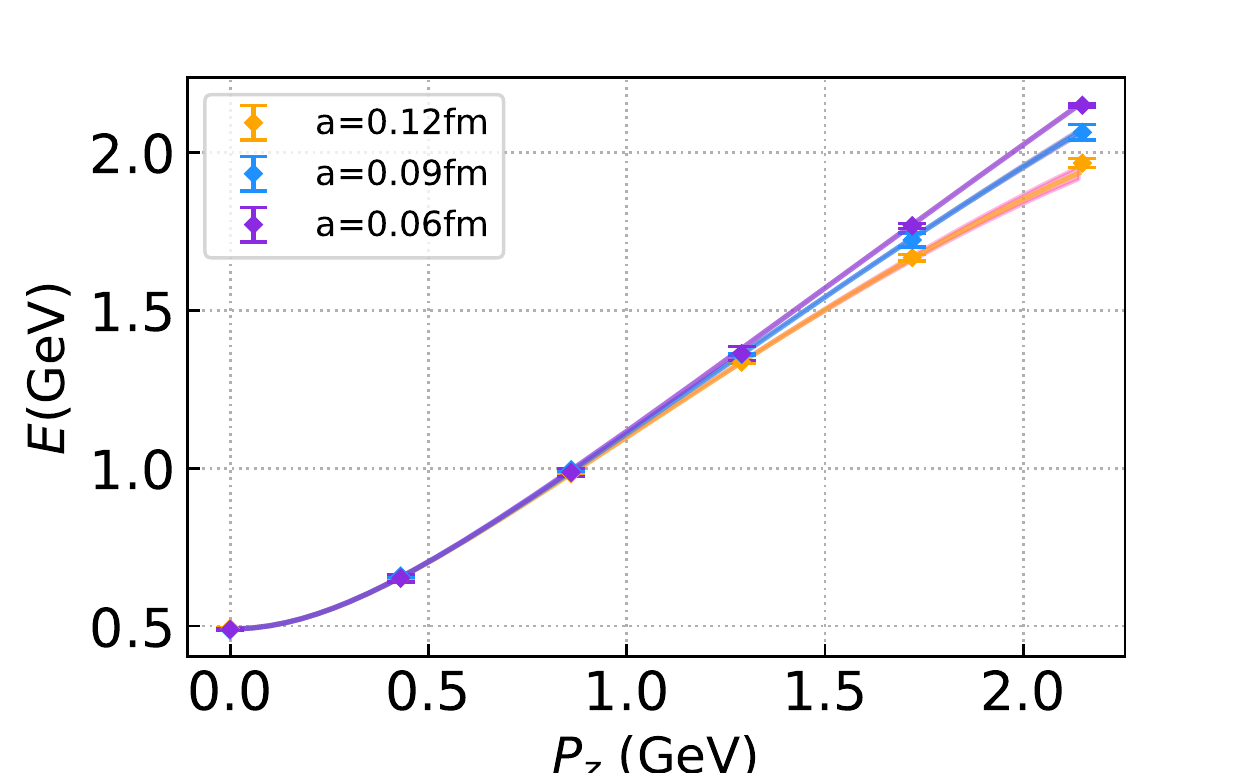}
\caption{Dispersion relation for the $\pi$ (top) and $K$ (bottom) meson.}\label{fig:dispersion_relation}
\end{center}
\end{figure}

\begin{figure}[!th]
\begin{center}
\includegraphics[width=0.47\textwidth]{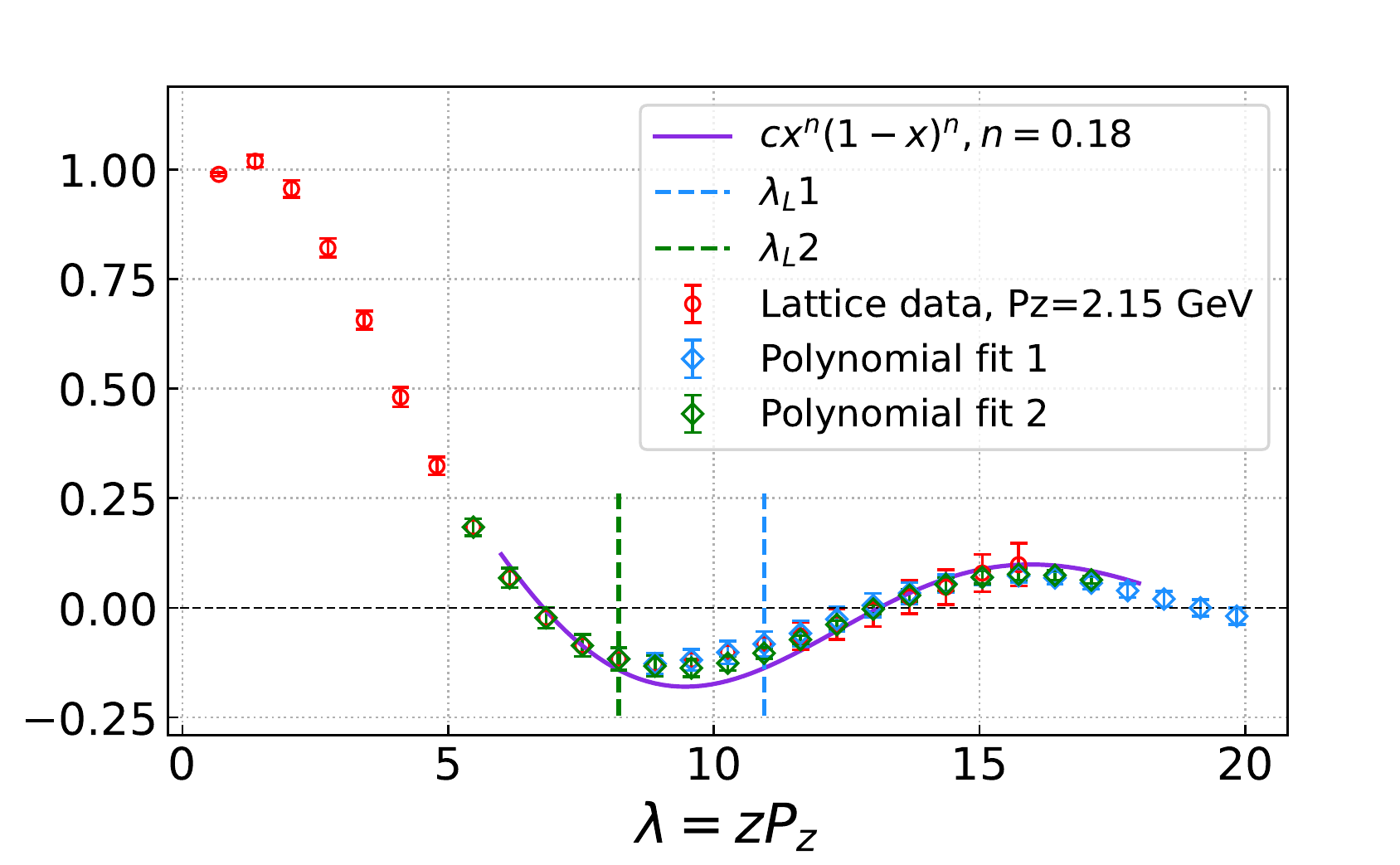}
\caption{The polynomial extrapolation of the pion quasi-LF correlation in the regions$(\lambda_{L1} >8.2)$ and $(\lambda_{L2} >11)$, for the case with $P_z=2.15$~GeV. }\label{fig:extrapolation}
\end{center}    
\end{figure}

\textit{Large $\lambda$ extrapolation:} In the hybrid scheme, we adopt a physics-based extrapolation~\cite{Ji:2020brr} at large quasi-LF distance ($\lambda=z P_z$). Depending on the hadron momentum, two extrapolation forms have been proposed:
\begin{align}
	  H^{\rm R}_{m}(z, P_z) &= \Big[\frac{c_1}{(i\lambda)^a} + e^{-i\lambda}\frac{c_2}{(-i \lambda)^b}\Big]e^{-\lambda/\lambda_0}, \nonumber\\
	  H^{\rm R}_{m}(z, P_z) &= \frac{c_1}{(i\lambda)^a} + e^{-i\lambda}\frac{c_2}{(-i \lambda)^b}.
	  \label{eq:sup_extra}
\end{align} 
where the exponential term comes from the expectation that at finite momentum the correlation function has a finite correlation length (denoted as $\lambda_0$), which goes to infinity when the momentum goes to infinity. In this work, for light $\pi, K$ meson at the physical point, the Lorentz boost factor is very large, so is the correlation length. We therefore use the formula in the second row in Eq.(\ref{eq:sup_extra}) for the extrapolation. For the pion, one can choose $c_1=c_2, \, a=b$ due to isospin symmetry, and fit them using available lattice results in the large $\lambda$ region. The pion result at $P_z=2.15$ GeV is shown in Fig.~\ref{fig:extrapolation}. We can see that the polynomial fit (blue/green diamonds) agrees with the original lattice data (red dots) well. The situation is similar in the case of the kaon. To estimate the modification effects of extrapolation form, we take two extrapolation regions($\lambda_{L1} >8.2$, $\lambda_{L2} >11$ in Fig.~\ref{fig:extrapolation} case) and treated their difference as an estimate of systematic error from extrapolation.

\begin{figure}[!th]
\begin{center}
\includegraphics[width=0.47\textwidth]{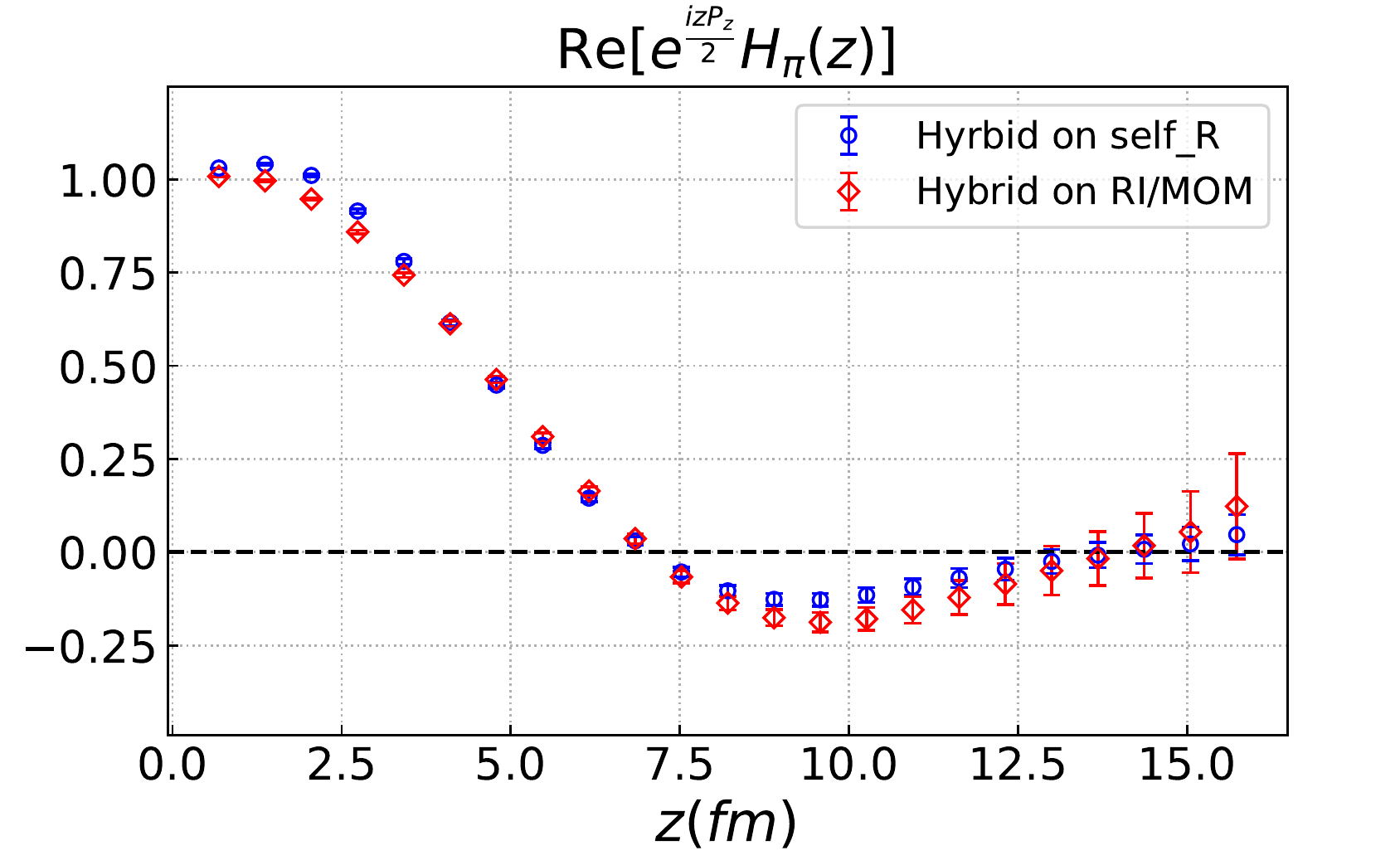}
\caption{ Comparison of renormalized DA matrix elements with $P_z=2.15$~GeV, $a=0.06$~fm for the hybrid scheme based on RI/MOM used in \cite{Hua:2020gnw} and based on self renormalization used in this work.}\label{fig:comp_2renorm}
\end{center}
\end{figure}

\textit{Renormaliation scheme dependence:} In this work, we have used a hybrid renormalization scheme based on self renormalization, where the renormalization factors are determined by fitting to the bare hadron matrix element. In a previous work~\cite{Hua:2020gnw}, a hybrid scheme based on RI/MOM was used, where the fitting was done for the off-shell quark matrix element. As the lattice spacings used there are not very small, the linear divergences appear to be canceled up to numerical uncertainty. However, the fitting form is not well justified theoretically. Nevertheless, here we make a comparison between the two schemes, and show the results in Fig.~\ref{fig:comp_2renorm}. 
As can be seen from the figure, the renormalized matrix elements in the two schemes deviate less than $1\sigma$ in most regions. Thus, the scheme from \cite{Hua:2020gnw} still yields consistent results with the hybrid scheme based on self renormalization, although the latter is theoretically more self-consistent.

\section{Observations from the data}
\begin{figure}[!th]
\begin{center}
\includegraphics[width=0.47\textwidth]{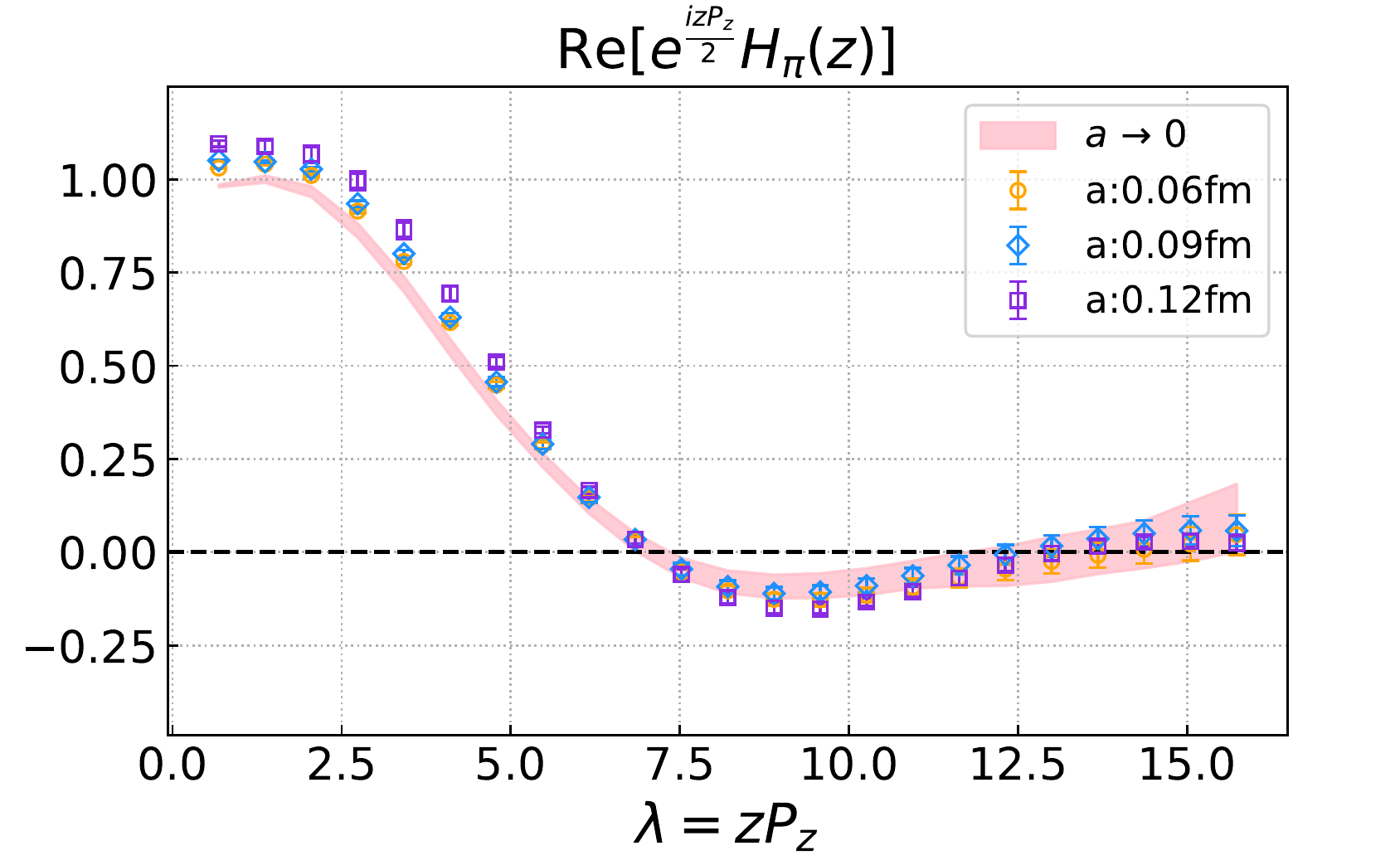}
\includegraphics[width=0.47\textwidth]{rotated_pion215_im.pdf}
\caption{The quasi-LF correlation function $e^{izP_z/2}H^{\rm R}_m(z)$ for the pion in coordinate space after the continuum limit $a \to 0$. We performed a phase rotation by a factor $e^{izP_z/2}$ with   $P_z=2.15$~GeV. }\label{fig:rota_matrix_pion}
\end{center}
\end{figure}

\textit{Phase rotation in coordinate space:}
We perform a phase rotation $e^{izP_z/2}$ for the renormalized $\pi, K$ correlations $H^{\rm R}_m(z,P_z)$, so that the imaginary part directly reflects the SU(3) flavor breaking effect. 
As shown in the real part (upper panel) of Fig.~\ref{fig:rota_matrix_pion}, the linear divergences have been canceled by the self renormalization, and the renormalized data for different lattice spacings are almost consistent with each other with only small residual discretization effects.

\begin{figure}[!th]
\begin{center}
\includegraphics[width=0.47\textwidth]{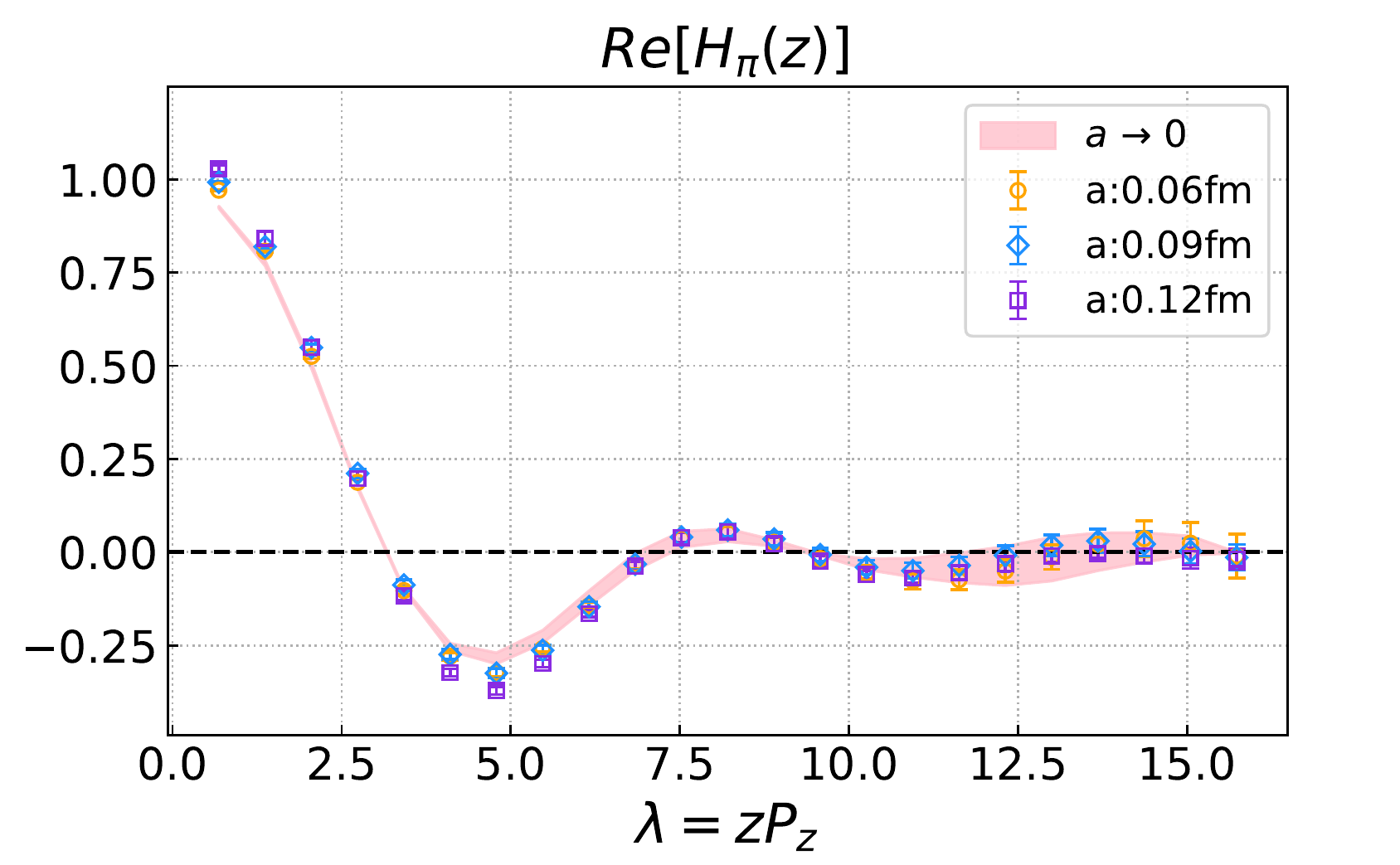}
\includegraphics[width=0.47\textwidth]{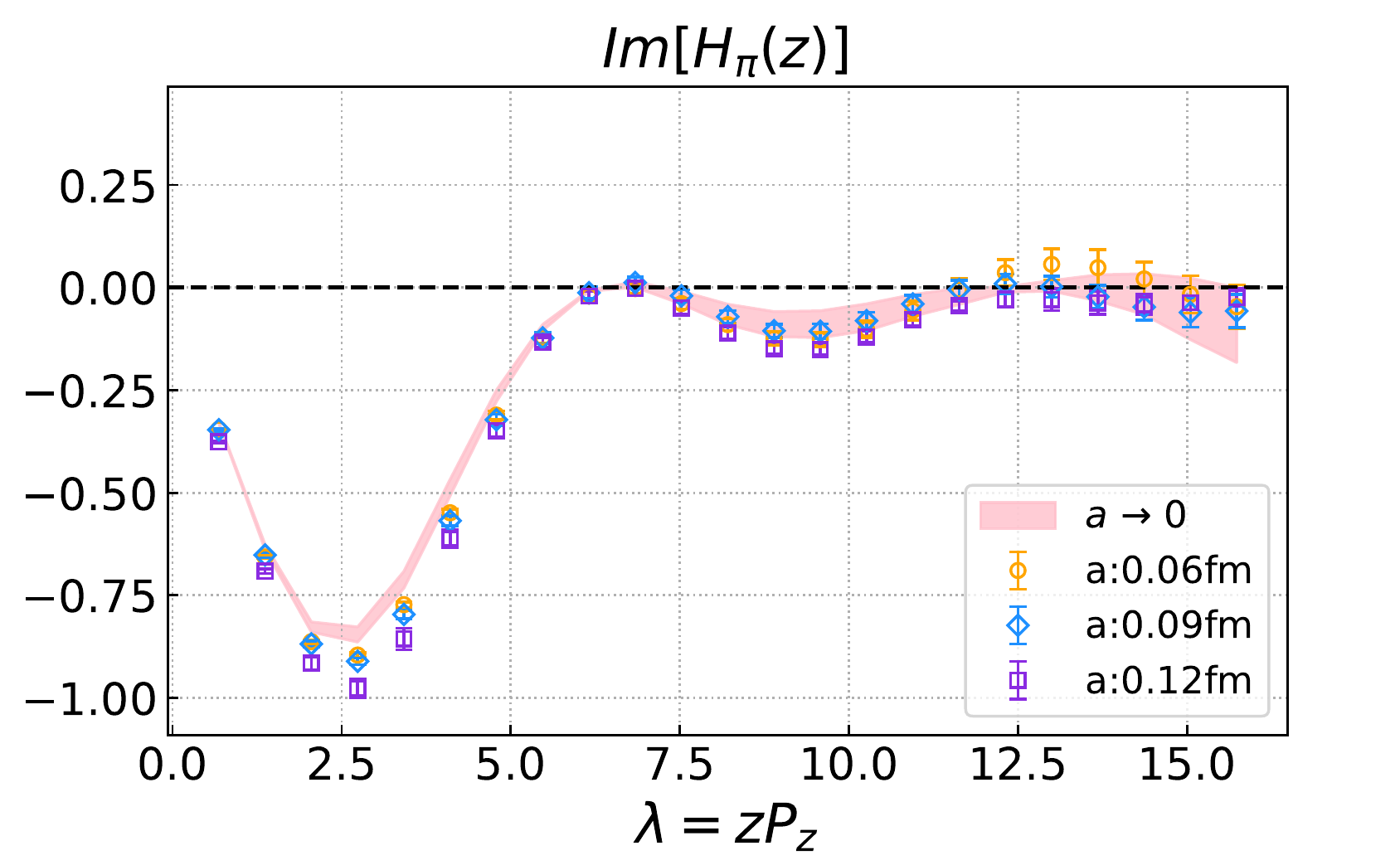}
\caption{The real part (top)  and imaginary part (bottom) of renormalized pion matrix elements $H^{\rm R}_m(z)$ for different lattice spacings at $P_z = 2.15$ GeV. }\label{fig:matrixks}
\end{center}
\end{figure}

For comparison purposes, we also show the original renormalized $\pi$ correlation without phase rotation in Fig.~\ref{fig:matrixks}. 

\begin{figure}[!th]
\begin{center}
\includegraphics[width=0.43\textwidth]{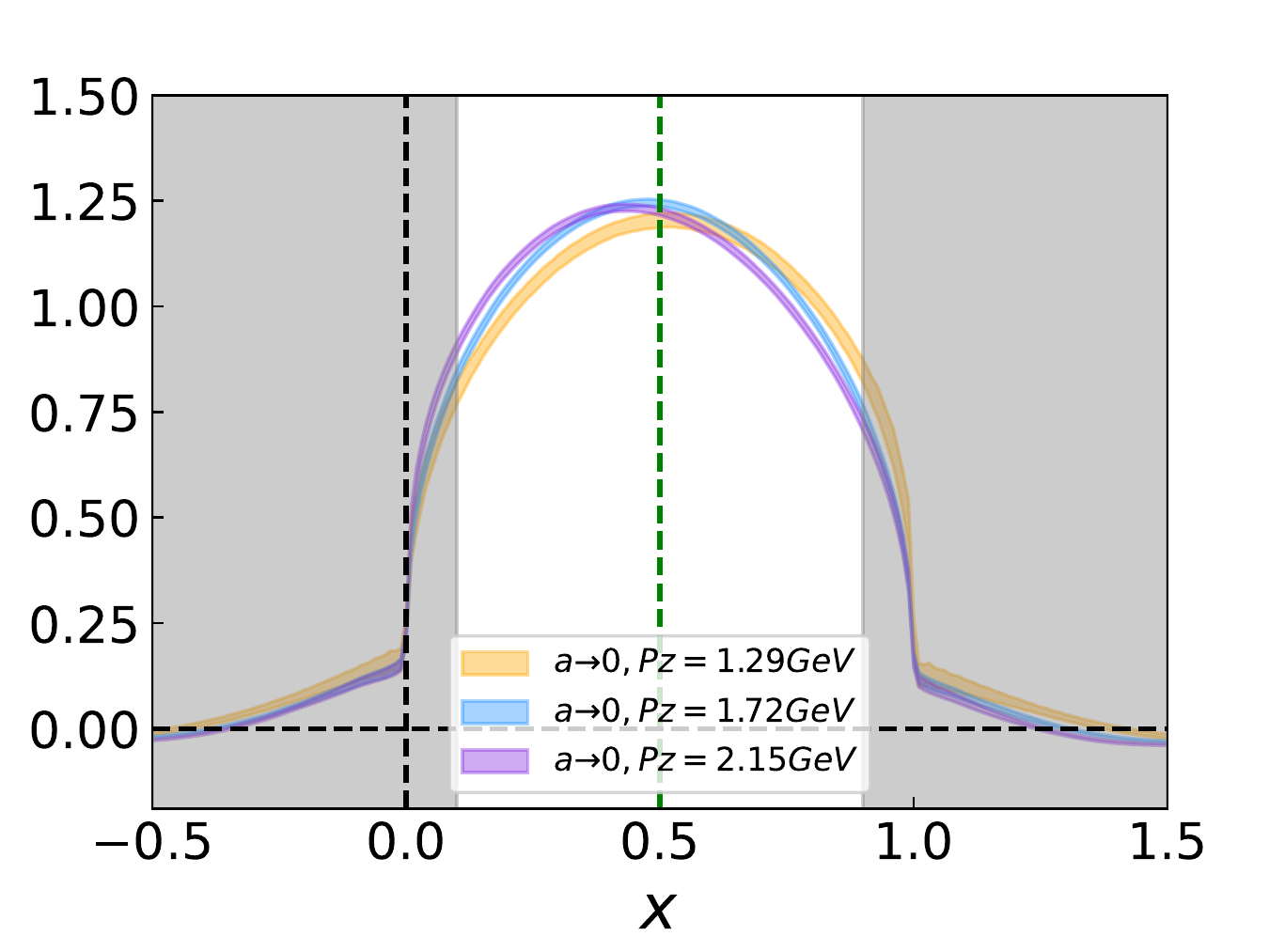}
\caption{Momentum dependence of continuum extrapolated results for the kaon.}\label{fig:pzlcda}
\end{center}
\end{figure}

\textit{$P_z$ dependence of Kaon DA in momentum space:} 
The momentum dependence of the continuum extrapolated results is shown in Fig.~\ref{fig:pzlcda} for the kaon. 
Since the endpoint region behavior cannot be reliably predicted by LaMET, we shade  the regions $x<0.1$ and $x>0.9$. 
As shown in the plot, the kaon DA is asymmetric around $x=1/2$ with the strange quark having on average a larger momentum fraction ($x$ is the momentum carried by the $u/d$ quark). This asymmetry increases slightly with $P_z$ in agreement with our expectation, since $s$ quark is heavier than $u/d$ quark. A similar asymmetry has also been observed in previous DA studies in LaMET \cite{Ji:2020brr,Zhang:2020gaj}. 


\textit{Prediction of the moments:} In phenomenological research, the DAs are usually expanded in terms of the Gegenbauer polynomials:
\begin{align}
	\phi(x) = 6x(1-x)\left[1 + \sum^\infty_{n}  a_nC_n^{3/2}(2x-1) \right] , 
\end{align}
where $a_n$s are Gegenbauer moments and $C_n$s are Gegenbauer polynomials. One can extract 
Gegenbauer moments from our results with this equation:
\begin{align}
	a_n = \frac{2(2n+3)}{3(n+1)(n+2)}\int^1_0 dx\, C_n^{3/2}(2x-1) \phi(x).
\end{align}
and the moments up to $a_4$ are given in Table.~\ref{tab:moment}. The odd moments of $\pi$ are identically zero due to the charge-conjugation invariance. The second moments from our calculation are $\xi_2^{\pi}=0.300(41)$, $\xi_2^{K}=0.258(32)$ ($\xi_n = \int^1_0 dx(2x-1)^n \phi(x)$), which agree better with a previous OPE calculation \cite{Braun:2006dg} rather than the recent one \cite{RQCD:2019osh}.

\begin{table*}[]
\caption{The Gegenbauer moments extracted from LCDAs, the numbers in the two brackets following the central value are the statistical error and the systematic error from the renormalization scale, large-$\lambda$ extrapolation, and continuum as well as infinite momentum extrapolation.}
\label{tab:moment}
\centering
\begin{tabular}{cccccccccc}
\hline
\hline
Gegenbauer moments &  $a_1$      &  $a_2$   &  $a_3$    &   $a_4$   \\
\hline
$\pi$       &  ~~ --- ~~ &  0.258(70)(52)   ~~~&   ~~ --- ~~&   0.122(46)(31) \\
$K$         &  -0.108(14)(51) ~~ &  0.170(14)(44)   ~~~&   -0.043(06)(22)~~~&   0.073(08)(21)  \\ 
\hline
\end{tabular}
\end{table*}

\begin{widetext}
\section{Matching in Coordinate Space}
The matching in coordinate space in the ${\overline{\rm MS}}$ scheme can be expressed as: 
\begin{eqnarray}
    H^{\overline{\rm MS}}_m\left(\lambda, P_{z}\right)=\int_{0}^{1} d x d y \theta(1-x-y) C\left(x, y, (\lambda/P_{z})^{2}, \mu \right) e^{-i y \lambda} h_m((1-x-y) \lambda, P_z),
\end{eqnarray}\

with
\begin{eqnarray}
    C\left(x, y, z^{2}, \mu\right)&=&\delta(x) \delta(y) 
    +\frac{\alpha_{s} C_{F}}{2 \pi}\left\{-\left(f\left(z^{2}, \mu^{2}\right)+1\right)\left[1+\left[\frac{1-x}{x}\right]_{+} \delta(y)+\left[\frac{1-y}{y}\right]_{+} \delta(x)-2 \delta(x) \delta(y)\right]\right. \\ \nonumber
    &&\left.+4-2\left[\frac{\ln x}{x}\right]_{+} \delta(y)-2\left[\frac{\ln y}{y}\right]_{+} \delta(x)\right\},
\end{eqnarray}
\begin{eqnarray}
    f\left(z^{2}, \mu^{2}\right)=\ln \left[\frac{z^{2} \mu^{2} \exp \left(2 \gamma_{E}\right)}{4}\right],
\end{eqnarray}
where $\mu$ is the factorization scale, $x,y$ are the Feynman integral parameters, $z$ is the non-local separation, $\lambda = zP_z$ is the quasi-LF distance for the quasi-DA. $h_m$ is the LF correlation related to the light-cone DA $\phi(u,\mu)$ through the following Fourier transformation ($h_m$ here has a different convention from $h^{R}_m$ in Eq. (10) of main text)
\begin{equation}
h_m((1-x-y)\lambda,P_{z})=\int_0^1 du\, e^{-i u (1-x-y) \lambda}\phi(u,\mu).
\end{equation}
Such a matching is related to the matching in momentum space~\cite{Ji:2015qla,Liu:2018tox} by Fourier transformation.

In practice, we extract the DA by applying an inverse matching on the quasi-DA in the ratio scheme, with the matching kernel being given by
\begin{eqnarray}
    C_{\text {ratio}}\left(x, y, z^{2}, \mu\right)=\frac{C\left(x, y, z^{2}, \mu\right)}{H^{\overline{\rm MS},\ {\rm 1-loop}}_m(z)}.
\end{eqnarray}
After some manipulations, the final matching in coordinate space can be expressed as:
\begin{eqnarray}
    H^{{\rm R}}_m\left(\lambda, P_{z}\right)&&=\int_{0}^{1} d x d y \theta(1-x-y) C_{\text {ratio}}\left(x, y, (\lambda/P_{z})^{2}, \mu \right) e^{-i y \lambda} h_m((1-x-y) \lambda, P_z)  \\ \nonumber
    &&= \int_{0}^{\lambda} d \lambda' M(\lambda, \lambda', \mu) h_m(\lambda', P_z) ,
\end{eqnarray}
where $\lambda'$ is the LF distance for the DA and:
\begin{eqnarray}
    M(\lambda, \lambda', \mu)  =&& \delta(\lambda - \lambda') + \frac{\alpha_s C_F}{2 \pi} \delta(\lambda - \lambda') \left[\frac{1}{2}f(z^2, \mu^2) - \frac{3}{2}\right] \\ \nonumber
    &&+ \frac{\alpha_s C_F}{2 \pi} \left[\frac{\lambda'}{\lambda - \lambda'}\right]_{+} \cdot \frac{1}{\lambda} \left[-(f(z^2, \mu^2)+1)(1+e^{-i(\lambda - \lambda')})\right] \\ \nonumber
    &&+ \frac{\alpha_s C_F}{2 \pi} \left[\frac{\ln(1-\frac{\lambda'}{\lambda})}{1-\frac{\lambda'}{\lambda}}\right]_{+} \cdot \frac{1}{\lambda} \left[-2(1 + e^{-i(\lambda - \lambda')})\right] + \frac{\alpha_s C_F}{2 \pi} \frac{1- e^{-i(\lambda - \lambda')}}{i \lambda^2} \left[3 - f(z^2, \mu^2)\right].
\end{eqnarray}
Fig.~\ref{fig:cor_cv_mom} shows a comparison between the coordinate and momentum space matching implementations. The results agree within errors in the physical region. The slight difference is due to different extrapolation choices,
as in the former case we can do the large $\lambda$ extrapolation after the matching, while in the latter the large $\lambda$ extrapolation has to be done before Fourier transform and matching.
The results from momentum space matching suffer from an oscillation at the endpoints which causes a more significant tail in the unphysical region. Such a behavior comes from artifacts related to the discontinuity of the momentum space matching kernel at the endpoints. 
\begin{figure}[!th]
\begin{center}
\includegraphics[width=0.47\textwidth]{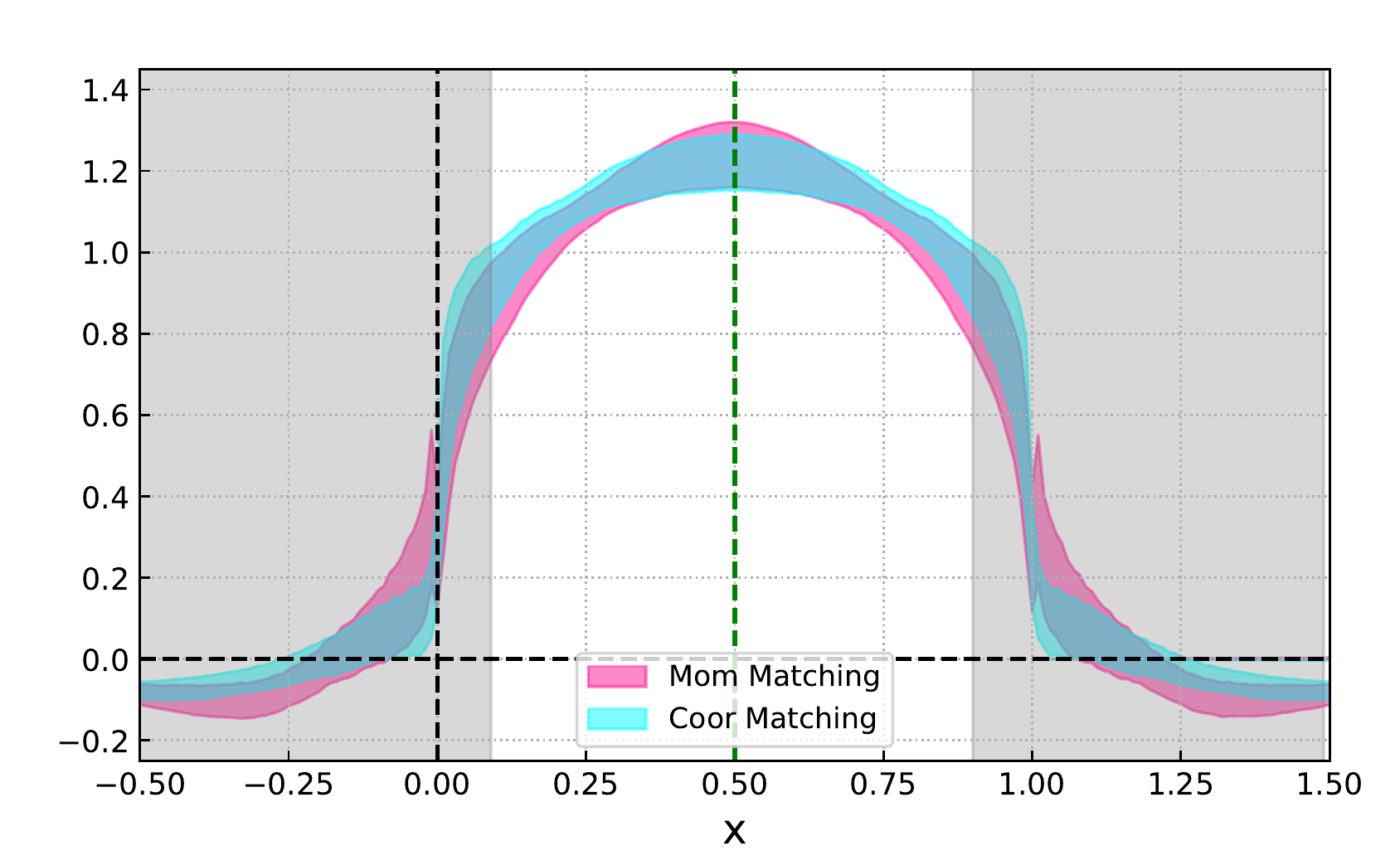}
\includegraphics[width=0.47\textwidth]{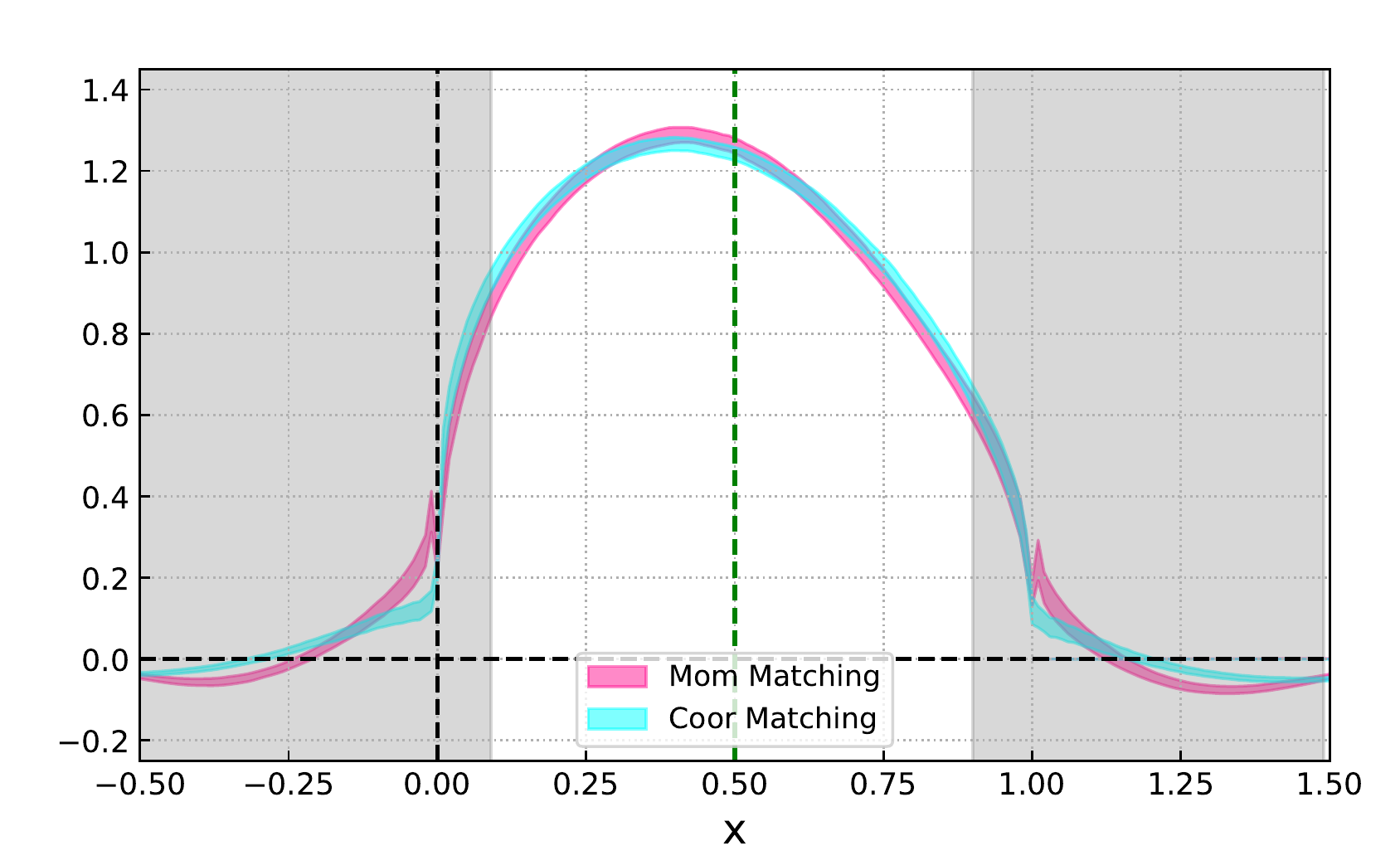}
\caption{ The comparison of matching in coordinate space and momentum space for the pion (left)  and kaon (right).}\label{fig:cor_cv_mom}
\end{center}
\end{figure}


\end{widetext}

\bibliography{ref}

\begin{thebibliography}{52}%
\makeatletter
\providecommand \@ifxundefined [1]{%
 \@ifx{#1\undefined}
}%
\providecommand \@ifnum [1]{%
 \ifnum #1\expandafter \@firstoftwo
 \else \expandafter \@secondoftwo
 \fi
}%
\providecommand \@ifx [1]{%
 \ifx #1\expandafter \@firstoftwo
 \else \expandafter \@secondoftwo
 \fi
}%
\providecommand \natexlab [1]{#1}%
\providecommand \enquote  [1]{``#1''}%
\providecommand \bibnamefont  [1]{#1}%
\providecommand \bibfnamefont [1]{#1}%
\providecommand \citenamefont [1]{#1}%
\providecommand \href@noop [0]{\@secondoftwo}%
\providecommand \href [0]{\begingroup \@sanitize@url \@href}%
\providecommand \@href[1]{\@@startlink{#1}\@@href}%
\providecommand \@@href[1]{\endgroup#1\@@endlink}%
\providecommand \@sanitize@url [0]{\catcode `\\12\catcode `\$12\catcode
  `\&12\catcode `\#12\catcode `\^12\catcode `\_12\catcode `\%12\relax}%
\providecommand \@@startlink[1]{}%
\providecommand \@@endlink[0]{}%
\providecommand \url  [0]{\begingroup\@sanitize@url \@url }%
\providecommand \@url [1]{\endgroup\@href {#1}{\urlprefix }}%
\providecommand \urlprefix  [0]{URL }%
\providecommand \Eprint [0]{\href }%
\providecommand \doibase [0]{http://dx.doi.org/}%
\providecommand \selectlanguage [0]{\@gobble}%
\providecommand \bibinfo  [0]{\@secondoftwo}%
\providecommand \bibfield  [0]{\@secondoftwo}%
\providecommand \translation [1]{[#1]}%
\providecommand \BibitemOpen [0]{}%
\providecommand \bibitemStop [0]{}%
\providecommand \bibitemNoStop [0]{.\EOS\space}%
\providecommand \EOS [0]{\spacefactor3000\relax}%
\providecommand \BibitemShut  [1]{\csname bibitem#1\endcsname}%
\let\auto@bib@innerbib\@empty
\bibitem [{\citenamefont {Weinberg}(2013)}]{Weinberg:1996kr}%
  \BibitemOpen
  \bibfield  {author} {\bibinfo {author} {\bibfnamefont {S.}~\bibnamefont
  {Weinberg}},\ }\href@noop {} {\emph {\bibinfo {title} {{The quantum theory of
  fields. Vol. 2: Modern applications}}}}\ (\bibinfo  {publisher} {Cambridge
  University Press},\ \bibinfo {year} {2013})\BibitemShut {NoStop}%
\bibitem [{\citenamefont {Cheng}\ and\ \citenamefont
  {Chua}(2009)}]{Cheng:2009cn}%
  \BibitemOpen
  \bibfield  {author} {\bibinfo {author} {\bibfnamefont {H.-Y.}\ \bibnamefont
  {Cheng}}\ and\ \bibinfo {author} {\bibfnamefont {C.-K.}\ \bibnamefont
  {Chua}},\ }\href {\doibase 10.1103/PhysRevD.80.114008} {\bibfield  {journal}
  {\bibinfo  {journal} {Phys. Rev. D}\ }\textbf {\bibinfo {volume} {80}},\
  \bibinfo {pages} {114008} (\bibinfo {year} {2009})},\ \Eprint
  {http://arxiv.org/abs/0909.5229} {arXiv:0909.5229 [hep-ph]} \BibitemShut
  {NoStop}%
\bibitem [{\citenamefont {Su}\ \emph {et~al.}(2011)\citenamefont {Su},
  \citenamefont {Wu}, \citenamefont {Yang},\ and\ \citenamefont
  {Zhuang}}]{Su:2010vt}%
  \BibitemOpen
  \bibfield  {author} {\bibinfo {author} {\bibfnamefont {F.}~\bibnamefont
  {Su}}, \bibinfo {author} {\bibfnamefont {Y.-L.}\ \bibnamefont {Wu}}, \bibinfo
  {author} {\bibfnamefont {Y.-B.}\ \bibnamefont {Yang}}, \ and\ \bibinfo
  {author} {\bibfnamefont {C.}~\bibnamefont {Zhuang}},\ }\href {\doibase
  10.1088/0954-3899/38/1/015006} {\bibfield  {journal} {\bibinfo  {journal} {J.
  Phys. G}\ }\textbf {\bibinfo {volume} {38}},\ \bibinfo {pages} {015006}
  (\bibinfo {year} {2011})},\ \Eprint {http://arxiv.org/abs/1006.1100}
  {arXiv:1006.1100 [hep-ph]} \BibitemShut {NoStop}%
\bibitem [{\citenamefont {Aaij}\ \emph {et~al.}(2019)\citenamefont {Aaij} \emph
  {et~al.}}]{LHCb:2019hip}%
  \BibitemOpen
  \bibfield  {author} {\bibinfo {author} {\bibfnamefont {R.}~\bibnamefont
  {Aaij}} \emph {et~al.} (\bibinfo {collaboration} {LHCb}),\ }\href {\doibase
  10.1103/PhysRevLett.122.191801} {\bibfield  {journal} {\bibinfo  {journal}
  {Phys. Rev. Lett.}\ }\textbf {\bibinfo {volume} {122}},\ \bibinfo {pages}
  {191801} (\bibinfo {year} {2019})},\ \Eprint
  {http://arxiv.org/abs/1903.09252} {arXiv:1903.09252 [hep-ex]} \BibitemShut
  {NoStop}%
\bibitem [{\citenamefont {Farrar}\ and\ \citenamefont
  {Jackson}(1979)}]{Farrar:1979aw}%
  \BibitemOpen
  \bibfield  {author} {\bibinfo {author} {\bibfnamefont {G.~R.}\ \bibnamefont
  {Farrar}}\ and\ \bibinfo {author} {\bibfnamefont {D.~R.}\ \bibnamefont
  {Jackson}},\ }\href {\doibase 10.1103/PhysRevLett.43.246} {\bibfield
  {journal} {\bibinfo  {journal} {Phys. Rev. Lett.}\ }\textbf {\bibinfo
  {volume} {43}},\ \bibinfo {pages} {246} (\bibinfo {year} {1979})}\BibitemShut
  {NoStop}%
\bibitem [{\citenamefont {Wang}\ and\ \citenamefont
  {Shen}(2017)}]{Wang:2017ijn}%
  \BibitemOpen
  \bibfield  {author} {\bibinfo {author} {\bibfnamefont {Y.-M.}\ \bibnamefont
  {Wang}}\ and\ \bibinfo {author} {\bibfnamefont {Y.-L.}\ \bibnamefont
  {Shen}},\ }\href {\doibase 10.1007/JHEP12(2017)037} {\bibfield  {journal}
  {\bibinfo  {journal} {JHEP}\ }\textbf {\bibinfo {volume} {12}},\ \bibinfo
  {pages} {037} (\bibinfo {year} {2017})},\ \Eprint
  {http://arxiv.org/abs/1706.05680} {arXiv:1706.05680 [hep-ph]} \BibitemShut
  {NoStop}%
\bibitem [{\citenamefont {Ji}(1997)}]{Ji:1996ek}%
  \BibitemOpen
  \bibfield  {author} {\bibinfo {author} {\bibfnamefont {X.-D.}\ \bibnamefont
  {Ji}},\ }\href {\doibase 10.1103/PhysRevLett.78.610} {\bibfield  {journal}
  {\bibinfo  {journal} {Phys. Rev. Lett.}\ }\textbf {\bibinfo {volume} {78}},\
  \bibinfo {pages} {610} (\bibinfo {year} {1997})},\ \Eprint
  {http://arxiv.org/abs/hep-ph/9603249} {arXiv:hep-ph/9603249} \BibitemShut
  {NoStop}%
\bibitem [{\citenamefont {Radyushkin}(1996)}]{Radyushkin:1996ru}%
  \BibitemOpen
  \bibfield  {author} {\bibinfo {author} {\bibfnamefont {A.~V.}\ \bibnamefont
  {Radyushkin}},\ }\href {\doibase 10.1016/0370-2693(96)00844-1} {\bibfield
  {journal} {\bibinfo  {journal} {Phys. Lett. B}\ }\textbf {\bibinfo {volume}
  {385}},\ \bibinfo {pages} {333} (\bibinfo {year} {1996})},\ \Eprint
  {http://arxiv.org/abs/hep-ph/9605431} {arXiv:hep-ph/9605431} \BibitemShut
  {NoStop}%
\bibitem [{\citenamefont {Lepage}\ and\ \citenamefont
  {Brodsky}(1979)}]{Lepage:1979zb}%
  \BibitemOpen
  \bibfield  {author} {\bibinfo {author} {\bibfnamefont {G.~P.}\ \bibnamefont
  {Lepage}}\ and\ \bibinfo {author} {\bibfnamefont {S.~J.}\ \bibnamefont
  {Brodsky}},\ }\href {\doibase 10.1016/0370-2693(79)90554-9} {\bibfield
  {journal} {\bibinfo  {journal} {Phys. Lett. B}\ }\textbf {\bibinfo {volume}
  {87}},\ \bibinfo {pages} {359} (\bibinfo {year} {1979})}\BibitemShut
  {NoStop}%
\bibitem [{\citenamefont {Chernyak}\ and\ \citenamefont
  {Zhitnitsky}(1982)}]{Chernyak:1981zz}%
  \BibitemOpen
  \bibfield  {author} {\bibinfo {author} {\bibfnamefont {V.~L.}\ \bibnamefont
  {Chernyak}}\ and\ \bibinfo {author} {\bibfnamefont {A.~R.}\ \bibnamefont
  {Zhitnitsky}},\ }\href {\doibase 10.1016/0550-3213(83)90251-1} {\bibfield
  {journal} {\bibinfo  {journal} {Nucl. Phys. B}\ }\textbf {\bibinfo {volume}
  {201}},\ \bibinfo {pages} {492} (\bibinfo {year} {1982})},\ \bibinfo {note}
  {[Erratum: Nucl.Phys.B 214, 547 (1983)]}\BibitemShut {NoStop}%
\bibitem [{\citenamefont {Ruiz~Arriola}\ and\ \citenamefont
  {Broniowski}(2006)}]{RuizArriola:2006jge}%
  \BibitemOpen
  \bibfield  {author} {\bibinfo {author} {\bibfnamefont {E.}~\bibnamefont
  {Ruiz~Arriola}}\ and\ \bibinfo {author} {\bibfnamefont {W.}~\bibnamefont
  {Broniowski}},\ }\href {\doibase 10.1103/PhysRevD.74.034008} {\bibfield
  {journal} {\bibinfo  {journal} {Phys. Rev. D}\ }\textbf {\bibinfo {volume}
  {74}},\ \bibinfo {pages} {034008} (\bibinfo {year} {2006})},\ \Eprint
  {http://arxiv.org/abs/hep-ph/0605318} {arXiv:hep-ph/0605318} \BibitemShut
  {NoStop}%
\bibitem [{\citenamefont {Radyushkin}(1994)}]{Radyushkin:1994xv}%
  \BibitemOpen
  \bibfield  {author} {\bibinfo {author} {\bibfnamefont {A.~V.}\ \bibnamefont
  {Radyushkin}},\ }in\ \href@noop {} {\emph {\bibinfo {booktitle} {{Workshop on
  Continuous Advances in QCD}}}}\ (\bibinfo {year} {1994})\ \Eprint
  {http://arxiv.org/abs/hep-ph/9406237} {arXiv:hep-ph/9406237} \BibitemShut
  {NoStop}%
\bibitem [{\citenamefont {Ruiz~Arriola}\ and\ \citenamefont
  {Broniowski}(2002)}]{RuizArriola:2002bp}%
  \BibitemOpen
  \bibfield  {author} {\bibinfo {author} {\bibfnamefont {E.}~\bibnamefont
  {Ruiz~Arriola}}\ and\ \bibinfo {author} {\bibfnamefont {W.}~\bibnamefont
  {Broniowski}},\ }\href {\doibase 10.1103/PhysRevD.66.094016} {\bibfield
  {journal} {\bibinfo  {journal} {Phys. Rev. D}\ }\textbf {\bibinfo {volume}
  {66}},\ \bibinfo {pages} {094016} (\bibinfo {year} {2002})},\ \Eprint
  {http://arxiv.org/abs/hep-ph/0207266} {arXiv:hep-ph/0207266} \BibitemShut
  {NoStop}%
\bibitem [{\citenamefont {Chang}\ \emph {et~al.}(2013)\citenamefont {Chang},
  \citenamefont {Cloet}, \citenamefont {Cobos-Martinez}, \citenamefont
  {Roberts}, \citenamefont {Schmidt},\ and\ \citenamefont
  {Tandy}}]{Chang:2013pq}%
  \BibitemOpen
  \bibfield  {author} {\bibinfo {author} {\bibfnamefont {L.}~\bibnamefont
  {Chang}}, \bibinfo {author} {\bibfnamefont {I.~C.}\ \bibnamefont {Cloet}},
  \bibinfo {author} {\bibfnamefont {J.~J.}\ \bibnamefont {Cobos-Martinez}},
  \bibinfo {author} {\bibfnamefont {C.~D.}\ \bibnamefont {Roberts}}, \bibinfo
  {author} {\bibfnamefont {S.~M.}\ \bibnamefont {Schmidt}}, \ and\ \bibinfo
  {author} {\bibfnamefont {P.~C.}\ \bibnamefont {Tandy}},\ }\href {\doibase
  10.1103/PhysRevLett.110.132001} {\bibfield  {journal} {\bibinfo  {journal}
  {Phys. Rev. Lett.}\ }\textbf {\bibinfo {volume} {110}},\ \bibinfo {pages}
  {132001} (\bibinfo {year} {2013})},\ \Eprint {http://arxiv.org/abs/1301.0324}
  {arXiv:1301.0324 [nucl-th]} \BibitemShut {NoStop}%
\bibitem [{\citenamefont {Agaev}\ \emph {et~al.}(2012)\citenamefont {Agaev},
  \citenamefont {Braun}, \citenamefont {Offen},\ and\ \citenamefont
  {Porkert}}]{Agaev:2012tm}%
  \BibitemOpen
  \bibfield  {author} {\bibinfo {author} {\bibfnamefont {S.~S.}\ \bibnamefont
  {Agaev}}, \bibinfo {author} {\bibfnamefont {V.~M.}\ \bibnamefont {Braun}},
  \bibinfo {author} {\bibfnamefont {N.}~\bibnamefont {Offen}}, \ and\ \bibinfo
  {author} {\bibfnamefont {F.~A.}\ \bibnamefont {Porkert}},\ }\href {\doibase
  10.1103/PhysRevD.86.077504} {\bibfield  {journal} {\bibinfo  {journal} {Phys.
  Rev. D}\ }\textbf {\bibinfo {volume} {86}},\ \bibinfo {pages} {077504}
  (\bibinfo {year} {2012})},\ \Eprint {http://arxiv.org/abs/1206.3968}
  {arXiv:1206.3968 [hep-ph]} \BibitemShut {NoStop}%
\bibitem [{\citenamefont {Shi}\ \emph {et~al.}(2014)\citenamefont {Shi},
  \citenamefont {Chang}, \citenamefont {Roberts}, \citenamefont {Schmidt},
  \citenamefont {Tandy},\ and\ \citenamefont {Zong}}]{Shi:2014uwa}%
  \BibitemOpen
  \bibfield  {author} {\bibinfo {author} {\bibfnamefont {C.}~\bibnamefont
  {Shi}}, \bibinfo {author} {\bibfnamefont {L.}~\bibnamefont {Chang}}, \bibinfo
  {author} {\bibfnamefont {C.~D.}\ \bibnamefont {Roberts}}, \bibinfo {author}
  {\bibfnamefont {S.~M.}\ \bibnamefont {Schmidt}}, \bibinfo {author}
  {\bibfnamefont {P.~C.}\ \bibnamefont {Tandy}}, \ and\ \bibinfo {author}
  {\bibfnamefont {H.-S.}\ \bibnamefont {Zong}},\ }\href {\doibase
  10.1016/j.physletb.2014.07.057} {\bibfield  {journal} {\bibinfo  {journal}
  {Phys. Lett. B}\ }\textbf {\bibinfo {volume} {738}},\ \bibinfo {pages} {512}
  (\bibinfo {year} {2014})},\ \Eprint {http://arxiv.org/abs/1406.3353}
  {arXiv:1406.3353 [nucl-th]} \BibitemShut {NoStop}%
\bibitem [{\citenamefont {Holt}\ and\ \citenamefont
  {Gilman}(2012)}]{Holt:2012gg}%
  \BibitemOpen
  \bibfield  {author} {\bibinfo {author} {\bibfnamefont {R.~J.}\ \bibnamefont
  {Holt}}\ and\ \bibinfo {author} {\bibfnamefont {R.}~\bibnamefont {Gilman}},\
  }\href {\doibase 10.1088/0034-4885/75/8/086301} {\bibfield  {journal}
  {\bibinfo  {journal} {Rept. Prog. Phys.}\ }\textbf {\bibinfo {volume} {75}},\
  \bibinfo {pages} {086301} (\bibinfo {year} {2012})},\ \Eprint
  {http://arxiv.org/abs/1205.5827} {arXiv:1205.5827 [nucl-ex]} \BibitemShut
  {NoStop}%
\bibitem [{\citenamefont {Goeckeler}\ \emph {et~al.}(2006)\citenamefont
  {Goeckeler}, \citenamefont {Horsley}, \citenamefont {Pleiter}, \citenamefont
  {Rakow}, \citenamefont {Schaefer}, \citenamefont {Schierholz}, \citenamefont
  {Schroers},\ and\ \citenamefont {Zanotti}}]{Gockeler:2005jz}%
  \BibitemOpen
  \bibfield  {author} {\bibinfo {author} {\bibfnamefont {M.}~\bibnamefont
  {Goeckeler}}, \bibinfo {author} {\bibfnamefont {R.}~\bibnamefont {Horsley}},
  \bibinfo {author} {\bibfnamefont {D.}~\bibnamefont {Pleiter}}, \bibinfo
  {author} {\bibfnamefont {P.~E.~L.}\ \bibnamefont {Rakow}}, \bibinfo {author}
  {\bibfnamefont {A.}~\bibnamefont {Schaefer}}, \bibinfo {author}
  {\bibfnamefont {G.}~\bibnamefont {Schierholz}}, \bibinfo {author}
  {\bibfnamefont {W.}~\bibnamefont {Schroers}}, \ and\ \bibinfo {author}
  {\bibfnamefont {J.~M.}\ \bibnamefont {Zanotti}},\ }\href {\doibase
  10.1016/j.nuclphysbps.2006.08.064} {\bibfield  {journal} {\bibinfo  {journal}
  {Nucl. Phys. B Proc. Suppl.}\ }\textbf {\bibinfo {volume} {161}},\ \bibinfo
  {pages} {69} (\bibinfo {year} {2006})},\ \Eprint
  {http://arxiv.org/abs/hep-lat/0510089} {arXiv:hep-lat/0510089} \BibitemShut
  {NoStop}%
\bibitem [{\citenamefont {Braun}\ \emph {et~al.}(2006)\citenamefont {Braun}
  \emph {et~al.}}]{Braun:2006dg}%
  \BibitemOpen
  \bibfield  {author} {\bibinfo {author} {\bibfnamefont {V.~M.}\ \bibnamefont
  {Braun}} \emph {et~al.},\ }\href {\doibase 10.1103/PhysRevD.74.074501}
  {\bibfield  {journal} {\bibinfo  {journal} {Phys. Rev. D}\ }\textbf {\bibinfo
  {volume} {74}},\ \bibinfo {pages} {074501} (\bibinfo {year} {2006})},\
  \Eprint {http://arxiv.org/abs/hep-lat/0606012} {arXiv:hep-lat/0606012}
  \BibitemShut {NoStop}%
\bibitem [{\citenamefont {Boyle}\ \emph {et~al.}(2006)\citenamefont {Boyle},
  \citenamefont {Donnellan}, \citenamefont {Flynn}, \citenamefont {Juttner},
  \citenamefont {Noaki}, \citenamefont {Sachrajda},\ and\ \citenamefont
  {Tweedie}}]{Boyle:2006pw}%
  \BibitemOpen
  \bibfield  {author} {\bibinfo {author} {\bibfnamefont {P.~A.}\ \bibnamefont
  {Boyle}}, \bibinfo {author} {\bibfnamefont {M.~A.}\ \bibnamefont
  {Donnellan}}, \bibinfo {author} {\bibfnamefont {J.~M.}\ \bibnamefont
  {Flynn}}, \bibinfo {author} {\bibfnamefont {A.}~\bibnamefont {Juttner}},
  \bibinfo {author} {\bibfnamefont {J.}~\bibnamefont {Noaki}}, \bibinfo
  {author} {\bibfnamefont {C.~T.}\ \bibnamefont {Sachrajda}}, \ and\ \bibinfo
  {author} {\bibfnamefont {R.~J.}\ \bibnamefont {Tweedie}} (\bibinfo
  {collaboration} {UKQCD}),\ }\href {\doibase 10.1016/j.physletb.2006.07.033}
  {\bibfield  {journal} {\bibinfo  {journal} {Phys. Lett. B}\ }\textbf
  {\bibinfo {volume} {641}},\ \bibinfo {pages} {67} (\bibinfo {year} {2006})},\
  \Eprint {http://arxiv.org/abs/hep-lat/0607018} {arXiv:hep-lat/0607018}
  \BibitemShut {NoStop}%
\bibitem [{\citenamefont {Arthur}\ \emph {et~al.}(2011)\citenamefont {Arthur},
  \citenamefont {Boyle}, \citenamefont {Brommel}, \citenamefont {Donnellan},
  \citenamefont {Flynn}, \citenamefont {Juttner}, \citenamefont {Rae},\ and\
  \citenamefont {Sachrajda}}]{Arthur:2010xf}%
  \BibitemOpen
  \bibfield  {author} {\bibinfo {author} {\bibfnamefont {R.}~\bibnamefont
  {Arthur}}, \bibinfo {author} {\bibfnamefont {P.~A.}\ \bibnamefont {Boyle}},
  \bibinfo {author} {\bibfnamefont {D.}~\bibnamefont {Brommel}}, \bibinfo
  {author} {\bibfnamefont {M.~A.}\ \bibnamefont {Donnellan}}, \bibinfo {author}
  {\bibfnamefont {J.~M.}\ \bibnamefont {Flynn}}, \bibinfo {author}
  {\bibfnamefont {A.}~\bibnamefont {Juttner}}, \bibinfo {author} {\bibfnamefont
  {T.~D.}\ \bibnamefont {Rae}}, \ and\ \bibinfo {author} {\bibfnamefont
  {C.~T.~C.}\ \bibnamefont {Sachrajda}},\ }\href {\doibase
  10.1103/PhysRevD.83.074505} {\bibfield  {journal} {\bibinfo  {journal} {Phys.
  Rev. D}\ }\textbf {\bibinfo {volume} {83}},\ \bibinfo {pages} {074505}
  (\bibinfo {year} {2011})},\ \Eprint {http://arxiv.org/abs/1011.5906}
  {arXiv:1011.5906 [hep-lat]} \BibitemShut {NoStop}%
\bibitem [{\citenamefont {Braun}\ \emph {et~al.}(2015)\citenamefont {Braun},
  \citenamefont {Collins}, \citenamefont {G\"ockeler}, \citenamefont
  {P\'erez-Rubio}, \citenamefont {Sch\"afer}, \citenamefont {Schiel},\ and\
  \citenamefont {Sternbeck}}]{Braun:2015axa}%
  \BibitemOpen
  \bibfield  {author} {\bibinfo {author} {\bibfnamefont {V.~M.}\ \bibnamefont
  {Braun}}, \bibinfo {author} {\bibfnamefont {S.}~\bibnamefont {Collins}},
  \bibinfo {author} {\bibfnamefont {M.}~\bibnamefont {G\"ockeler}}, \bibinfo
  {author} {\bibfnamefont {P.}~\bibnamefont {P\'erez-Rubio}}, \bibinfo {author}
  {\bibfnamefont {A.}~\bibnamefont {Sch\"afer}}, \bibinfo {author}
  {\bibfnamefont {R.~W.}\ \bibnamefont {Schiel}}, \ and\ \bibinfo {author}
  {\bibfnamefont {A.}~\bibnamefont {Sternbeck}},\ }\href {\doibase
  10.1103/PhysRevD.92.014504} {\bibfield  {journal} {\bibinfo  {journal} {Phys.
  Rev. D}\ }\textbf {\bibinfo {volume} {92}},\ \bibinfo {pages} {014504}
  (\bibinfo {year} {2015})},\ \Eprint {http://arxiv.org/abs/1503.03656}
  {arXiv:1503.03656 [hep-lat]} \BibitemShut {NoStop}%
\bibitem [{\citenamefont {Bali}\ \emph {et~al.}(2017)\citenamefont {Bali},
  \citenamefont {Braun}, \citenamefont {G\"ockeler}, \citenamefont {Gruber},
  \citenamefont {Hutzler}, \citenamefont {Korcyl}, \citenamefont {Lang},\ and\
  \citenamefont {Sch\"afer}}]{Bali:2017ude}%
  \BibitemOpen
  \bibfield  {author} {\bibinfo {author} {\bibfnamefont {G.~S.}\ \bibnamefont
  {Bali}}, \bibinfo {author} {\bibfnamefont {V.~M.}\ \bibnamefont {Braun}},
  \bibinfo {author} {\bibfnamefont {M.}~\bibnamefont {G\"ockeler}}, \bibinfo
  {author} {\bibfnamefont {M.}~\bibnamefont {Gruber}}, \bibinfo {author}
  {\bibfnamefont {F.}~\bibnamefont {Hutzler}}, \bibinfo {author} {\bibfnamefont
  {P.}~\bibnamefont {Korcyl}}, \bibinfo {author} {\bibfnamefont
  {B.}~\bibnamefont {Lang}}, \ and\ \bibinfo {author} {\bibfnamefont
  {A.}~\bibnamefont {Sch\"afer}} (\bibinfo {collaboration} {RQCD}),\ }\href
  {\doibase 10.1016/j.physletb.2017.08.077} {\bibfield  {journal} {\bibinfo
  {journal} {Phys. Lett. B}\ }\textbf {\bibinfo {volume} {774}},\ \bibinfo
  {pages} {91} (\bibinfo {year} {2017})},\ \Eprint
  {http://arxiv.org/abs/1705.10236} {arXiv:1705.10236 [hep-lat]} \BibitemShut
  {NoStop}%
\bibitem [{\citenamefont {Bali}\ \emph {et~al.}(2019)\citenamefont {Bali},
  \citenamefont {Braun}, \citenamefont {B\"urger}, \citenamefont {G\"ockeler},
  \citenamefont {Gruber}, \citenamefont {Hutzler}, \citenamefont {Korcyl},
  \citenamefont {Sch\"afer}, \citenamefont {Sternbeck},\ and\ \citenamefont
  {Wein}}]{RQCD:2019osh}%
  \BibitemOpen
  \bibfield  {author} {\bibinfo {author} {\bibfnamefont {G.~S.}\ \bibnamefont
  {Bali}}, \bibinfo {author} {\bibfnamefont {V.~M.}\ \bibnamefont {Braun}},
  \bibinfo {author} {\bibfnamefont {S.}~\bibnamefont {B\"urger}}, \bibinfo
  {author} {\bibfnamefont {M.}~\bibnamefont {G\"ockeler}}, \bibinfo {author}
  {\bibfnamefont {M.}~\bibnamefont {Gruber}}, \bibinfo {author} {\bibfnamefont
  {F.}~\bibnamefont {Hutzler}}, \bibinfo {author} {\bibfnamefont
  {P.}~\bibnamefont {Korcyl}}, \bibinfo {author} {\bibfnamefont
  {A.}~\bibnamefont {Sch\"afer}}, \bibinfo {author} {\bibfnamefont
  {A.}~\bibnamefont {Sternbeck}}, \ and\ \bibinfo {author} {\bibfnamefont
  {P.}~\bibnamefont {Wein}} (\bibinfo {collaboration} {RQCD}),\ }\href
  {\doibase 10.1007/JHEP08(2019)065} {\bibfield  {journal} {\bibinfo  {journal}
  {JHEP}\ }\textbf {\bibinfo {volume} {08}},\ \bibinfo {pages} {065} (\bibinfo
  {year} {2019})},\ \bibinfo {note} {[Addendum: JHEP 11, 037 (2020)]},\ \Eprint
  {http://arxiv.org/abs/1903.08038} {arXiv:1903.08038 [hep-lat]} \BibitemShut
  {NoStop}%
\bibitem [{\citenamefont {Ji}(2013)}]{Ji:2013dva}%
  \BibitemOpen
  \bibfield  {author} {\bibinfo {author} {\bibfnamefont {X.}~\bibnamefont
  {Ji}},\ }\href {\doibase 10.1103/PhysRevLett.110.262002} {\bibfield
  {journal} {\bibinfo  {journal} {Phys. Rev. Lett.}\ }\textbf {\bibinfo
  {volume} {110}},\ \bibinfo {pages} {262002} (\bibinfo {year} {2013})},\
  \Eprint {http://arxiv.org/abs/1305.1539} {arXiv:1305.1539 [hep-ph]}
  \BibitemShut {NoStop}%
\bibitem [{\citenamefont {Ji}(2014)}]{Ji:2014gla}%
  \BibitemOpen
  \bibfield  {author} {\bibinfo {author} {\bibfnamefont {X.}~\bibnamefont
  {Ji}},\ }\href {\doibase 10.1007/s11433-014-5492-3} {\bibfield  {journal}
  {\bibinfo  {journal} {Sci. China Phys. Mech. Astron.}\ }\textbf {\bibinfo
  {volume} {57}},\ \bibinfo {pages} {1407} (\bibinfo {year} {2014})},\ \Eprint
  {http://arxiv.org/abs/1404.6680} {arXiv:1404.6680 [hep-ph]} \BibitemShut
  {NoStop}%
\bibitem [{\citenamefont {Ji}\ \emph {et~al.}(2021{\natexlab{a}})\citenamefont
  {Ji}, \citenamefont {Liu}, \citenamefont {Liu}, \citenamefont {Zhang},\ and\
  \citenamefont {Zhao}}]{Ji:2020ect}%
  \BibitemOpen
  \bibfield  {author} {\bibinfo {author} {\bibfnamefont {X.}~\bibnamefont
  {Ji}}, \bibinfo {author} {\bibfnamefont {Y.-S.}\ \bibnamefont {Liu}},
  \bibinfo {author} {\bibfnamefont {Y.}~\bibnamefont {Liu}}, \bibinfo {author}
  {\bibfnamefont {J.-H.}\ \bibnamefont {Zhang}}, \ and\ \bibinfo {author}
  {\bibfnamefont {Y.}~\bibnamefont {Zhao}},\ }\href {\doibase
  10.1103/RevModPhys.93.035005} {\bibfield  {journal} {\bibinfo  {journal}
  {Rev. Mod. Phys.}\ }\textbf {\bibinfo {volume} {93}},\ \bibinfo {pages}
  {035005} (\bibinfo {year} {2021}{\natexlab{a}})},\ \Eprint
  {http://arxiv.org/abs/2004.03543} {arXiv:2004.03543 [hep-ph]} \BibitemShut
  {NoStop}%
\bibitem [{\citenamefont {Braun}\ and\ \citenamefont
  {M\"uller}(2008)}]{Braun:2007wv}%
  \BibitemOpen
  \bibfield  {author} {\bibinfo {author} {\bibfnamefont {V.}~\bibnamefont
  {Braun}}\ and\ \bibinfo {author} {\bibfnamefont {D.}~\bibnamefont
  {M\"uller}},\ }\href {\doibase 10.1140/epjc/s10052-008-0608-4} {\bibfield
  {journal} {\bibinfo  {journal} {Eur. Phys. J. C}\ }\textbf {\bibinfo {volume}
  {55}},\ \bibinfo {pages} {349} (\bibinfo {year} {2008})},\ \Eprint
  {http://arxiv.org/abs/0709.1348} {arXiv:0709.1348 [hep-ph]} \BibitemShut
  {NoStop}%
\bibitem [{\citenamefont {Bali}\ \emph
  {et~al.}(2018{\natexlab{a}})\citenamefont {Bali} \emph
  {et~al.}}]{Bali:2017gfr}%
  \BibitemOpen
  \bibfield  {author} {\bibinfo {author} {\bibfnamefont {G.~S.}\ \bibnamefont
  {Bali}} \emph {et~al.},\ }\href {\doibase 10.1140/epjc/s10052-018-5700-9}
  {\bibfield  {journal} {\bibinfo  {journal} {Eur. Phys. J. C}\ }\textbf
  {\bibinfo {volume} {78}},\ \bibinfo {pages} {217} (\bibinfo {year}
  {2018}{\natexlab{a}})},\ \Eprint {http://arxiv.org/abs/1709.04325}
  {arXiv:1709.04325 [hep-lat]} \BibitemShut {NoStop}%
\bibitem [{\citenamefont {Bali}\ \emph
  {et~al.}(2018{\natexlab{b}})\citenamefont {Bali}, \citenamefont {Braun},
  \citenamefont {Gl\"a\ss{}le}, \citenamefont {G\"ockeler}, \citenamefont
  {Gruber}, \citenamefont {Hutzler}, \citenamefont {Korcyl}, \citenamefont
  {Sch\"afer}, \citenamefont {Wein},\ and\ \citenamefont
  {Zhang}}]{Bali:2018spj}%
  \BibitemOpen
  \bibfield  {author} {\bibinfo {author} {\bibfnamefont {G.~S.}\ \bibnamefont
  {Bali}}, \bibinfo {author} {\bibfnamefont {V.~M.}\ \bibnamefont {Braun}},
  \bibinfo {author} {\bibfnamefont {B.}~\bibnamefont {Gl\"a\ss{}le}}, \bibinfo
  {author} {\bibfnamefont {M.}~\bibnamefont {G\"ockeler}}, \bibinfo {author}
  {\bibfnamefont {M.}~\bibnamefont {Gruber}}, \bibinfo {author} {\bibfnamefont
  {F.}~\bibnamefont {Hutzler}}, \bibinfo {author} {\bibfnamefont
  {P.}~\bibnamefont {Korcyl}}, \bibinfo {author} {\bibfnamefont
  {A.}~\bibnamefont {Sch\"afer}}, \bibinfo {author} {\bibfnamefont
  {P.}~\bibnamefont {Wein}}, \ and\ \bibinfo {author} {\bibfnamefont {J.-H.}\
  \bibnamefont {Zhang}},\ }\href {\doibase 10.1103/PhysRevD.98.094507}
  {\bibfield  {journal} {\bibinfo  {journal} {Phys. Rev. D}\ }\textbf {\bibinfo
  {volume} {98}},\ \bibinfo {pages} {094507} (\bibinfo {year}
  {2018}{\natexlab{b}})},\ \Eprint {http://arxiv.org/abs/1807.06671}
  {arXiv:1807.06671 [hep-lat]} \BibitemShut {NoStop}%
\bibitem [{\citenamefont {Detmold}\ \emph {et~al.}(2021)\citenamefont
  {Detmold}, \citenamefont {Grebe}, \citenamefont {Kanamori}, \citenamefont
  {Lin}, \citenamefont {Mondal}, \citenamefont {Perry},\ and\ \citenamefont
  {Zhao}}]{Detmold:2021qln}%
  \BibitemOpen
  \bibfield  {author} {\bibinfo {author} {\bibfnamefont {W.}~\bibnamefont
  {Detmold}}, \bibinfo {author} {\bibfnamefont {A.}~\bibnamefont {Grebe}},
  \bibinfo {author} {\bibfnamefont {I.}~\bibnamefont {Kanamori}}, \bibinfo
  {author} {\bibfnamefont {C.~J.~D.}\ \bibnamefont {Lin}}, \bibinfo {author}
  {\bibfnamefont {S.}~\bibnamefont {Mondal}}, \bibinfo {author} {\bibfnamefont
  {R.}~\bibnamefont {Perry}}, \ and\ \bibinfo {author} {\bibfnamefont
  {Y.}~\bibnamefont {Zhao}},\ }\href@noop {} {\  (\bibinfo {year} {2021})},\
  \Eprint {http://arxiv.org/abs/2109.15241} {arXiv:2109.15241 [hep-lat]}
  \BibitemShut {NoStop}%
\bibitem [{\citenamefont {Zhang}\ \emph {et~al.}(2017)\citenamefont {Zhang},
  \citenamefont {Chen}, \citenamefont {Ji}, \citenamefont {Jin},\ and\
  \citenamefont {Lin}}]{Zhang:2017bzy}%
  \BibitemOpen
  \bibfield  {author} {\bibinfo {author} {\bibfnamefont {J.-H.}\ \bibnamefont
  {Zhang}}, \bibinfo {author} {\bibfnamefont {J.-W.}\ \bibnamefont {Chen}},
  \bibinfo {author} {\bibfnamefont {X.}~\bibnamefont {Ji}}, \bibinfo {author}
  {\bibfnamefont {L.}~\bibnamefont {Jin}}, \ and\ \bibinfo {author}
  {\bibfnamefont {H.-W.}\ \bibnamefont {Lin}},\ }\href {\doibase
  10.1103/PhysRevD.95.094514} {\bibfield  {journal} {\bibinfo  {journal} {Phys.
  Rev. D}\ }\textbf {\bibinfo {volume} {95}},\ \bibinfo {pages} {094514}
  (\bibinfo {year} {2017})},\ \Eprint {http://arxiv.org/abs/1702.00008}
  {arXiv:1702.00008 [hep-lat]} \BibitemShut {NoStop}%
\bibitem [{\citenamefont {Zhang}\ \emph {et~al.}(2019)\citenamefont {Zhang},
  \citenamefont {Jin}, \citenamefont {Lin}, \citenamefont {Sch\"afer},
  \citenamefont {Sun}, \citenamefont {Yang}, \citenamefont {Zhang},
  \citenamefont {Zhao},\ and\ \citenamefont {Chen}}]{Zhang:2017zfe}%
  \BibitemOpen
  \bibfield  {author} {\bibinfo {author} {\bibfnamefont {J.-H.}\ \bibnamefont
  {Zhang}}, \bibinfo {author} {\bibfnamefont {L.}~\bibnamefont {Jin}}, \bibinfo
  {author} {\bibfnamefont {H.-W.}\ \bibnamefont {Lin}}, \bibinfo {author}
  {\bibfnamefont {A.}~\bibnamefont {Sch\"afer}}, \bibinfo {author}
  {\bibfnamefont {P.}~\bibnamefont {Sun}}, \bibinfo {author} {\bibfnamefont
  {Y.-B.}\ \bibnamefont {Yang}}, \bibinfo {author} {\bibfnamefont
  {R.}~\bibnamefont {Zhang}}, \bibinfo {author} {\bibfnamefont
  {Y.}~\bibnamefont {Zhao}}, \ and\ \bibinfo {author} {\bibfnamefont {J.-W.}\
  \bibnamefont {Chen}} (\bibinfo {collaboration} {LP3}),\ }\href {\doibase
  10.1016/j.nuclphysb.2018.12.020} {\bibfield  {journal} {\bibinfo  {journal}
  {Nucl. Phys. B}\ }\textbf {\bibinfo {volume} {939}},\ \bibinfo {pages} {429}
  (\bibinfo {year} {2019})},\ \Eprint {http://arxiv.org/abs/1712.10025}
  {arXiv:1712.10025 [hep-ph]} \BibitemShut {NoStop}%
\bibitem [{\citenamefont {Zhang}\ \emph {et~al.}(2020)\citenamefont {Zhang},
  \citenamefont {Honkala}, \citenamefont {Lin},\ and\ \citenamefont
  {Chen}}]{Zhang:2020gaj}%
  \BibitemOpen
  \bibfield  {author} {\bibinfo {author} {\bibfnamefont {R.}~\bibnamefont
  {Zhang}}, \bibinfo {author} {\bibfnamefont {C.}~\bibnamefont {Honkala}},
  \bibinfo {author} {\bibfnamefont {H.-W.}\ \bibnamefont {Lin}}, \ and\
  \bibinfo {author} {\bibfnamefont {J.-W.}\ \bibnamefont {Chen}},\ }\href
  {\doibase 10.1103/PhysRevD.102.094519} {\bibfield  {journal} {\bibinfo
  {journal} {Phys. Rev. D}\ }\textbf {\bibinfo {volume} {102}},\ \bibinfo
  {pages} {094519} (\bibinfo {year} {2020})},\ \Eprint
  {http://arxiv.org/abs/2005.13955} {arXiv:2005.13955 [hep-lat]} \BibitemShut
  {NoStop}%
\bibitem [{\citenamefont {Zhang}\ \emph {et~al.}(2021)\citenamefont {Zhang},
  \citenamefont {Li}, \citenamefont {Huo}, \citenamefont {Sch\"afer},
  \citenamefont {Sun},\ and\ \citenamefont {Yang}}]{Zhang:2020rsx}%
  \BibitemOpen
  \bibfield  {author} {\bibinfo {author} {\bibfnamefont {K.}~\bibnamefont
  {Zhang}}, \bibinfo {author} {\bibfnamefont {Y.-Y.}\ \bibnamefont {Li}},
  \bibinfo {author} {\bibfnamefont {Y.-K.}\ \bibnamefont {Huo}}, \bibinfo
  {author} {\bibfnamefont {A.}~\bibnamefont {Sch\"afer}}, \bibinfo {author}
  {\bibfnamefont {P.}~\bibnamefont {Sun}}, \ and\ \bibinfo {author}
  {\bibfnamefont {Y.-B.}\ \bibnamefont {Yang}} (\bibinfo {collaboration}
  {\ensuremath{\chi}QCD}),\ }\href {\doibase 10.1103/PhysRevD.104.074501}
  {\bibfield  {journal} {\bibinfo  {journal} {Phys. Rev. D}\ }\textbf {\bibinfo
  {volume} {104}},\ \bibinfo {pages} {074501} (\bibinfo {year} {2021})},\
  \Eprint {http://arxiv.org/abs/2012.05448} {arXiv:2012.05448 [hep-lat]}
  \BibitemShut {NoStop}%
\bibitem [{\citenamefont {Huo}\ \emph {et~al.}(2021)\citenamefont {Huo} \emph
  {et~al.}}]{LPC:2021xdx}%
  \BibitemOpen
  \bibfield  {author} {\bibinfo {author} {\bibfnamefont {Y.-K.}\ \bibnamefont
  {Huo}} \emph {et~al.} (\bibinfo {collaboration} {Lattice Parton Collaboration
  (LPC)}),\ }\href {\doibase 10.1016/j.nuclphysb.2021.115443} {\bibfield
  {journal} {\bibinfo  {journal} {Nucl. Phys. B}\ }\textbf {\bibinfo {volume}
  {969}},\ \bibinfo {pages} {115443} (\bibinfo {year} {2021})},\ \Eprint
  {http://arxiv.org/abs/2103.02965} {arXiv:2103.02965 [hep-lat]} \BibitemShut
  {NoStop}%
\bibitem [{sup()}]{supplemental}%
  \BibitemOpen
  \href@noop {} {\bibinfo  {journal} {Supplemental materials}\ }\BibitemShut
  {NoStop}%
\bibitem [{\citenamefont {Follana}\ \emph {et~al.}(2007)\citenamefont
  {Follana}, \citenamefont {Mason}, \citenamefont {Davies}, \citenamefont
  {Hornbostel}, \citenamefont {Lepage}, \citenamefont {Shigemitsu},
  \citenamefont {Trottier},\ and\ \citenamefont {Wong}}]{Follana:2006rc}%
  \BibitemOpen
\bibfield  {journal} {  }\bibfield  {author} {\bibinfo {author} {\bibfnamefont
  {E.}~\bibnamefont {Follana}}, \bibinfo {author} {\bibfnamefont
  {Q.}~\bibnamefont {Mason}}, \bibinfo {author} {\bibfnamefont
  {C.}~\bibnamefont {Davies}}, \bibinfo {author} {\bibfnamefont
  {K.}~\bibnamefont {Hornbostel}}, \bibinfo {author} {\bibfnamefont {G.~P.}\
  \bibnamefont {Lepage}}, \bibinfo {author} {\bibfnamefont {J.}~\bibnamefont
  {Shigemitsu}}, \bibinfo {author} {\bibfnamefont {H.}~\bibnamefont
  {Trottier}}, \ and\ \bibinfo {author} {\bibfnamefont {K.}~\bibnamefont
  {Wong}} (\bibinfo {collaboration} {HPQCD, UKQCD}),\ }\href {\doibase
  10.1103/PhysRevD.75.054502} {\bibfield  {journal} {\bibinfo  {journal} {Phys.
  Rev. D}\ }\textbf {\bibinfo {volume} {75}},\ \bibinfo {pages} {054502}
  (\bibinfo {year} {2007})},\ \Eprint {http://arxiv.org/abs/hep-lat/0610092}
  {arXiv:hep-lat/0610092} \BibitemShut {NoStop}%
\bibitem [{\citenamefont {Bazavov}\ \emph {et~al.}(2013)\citenamefont {Bazavov}
  \emph {et~al.}}]{MILC:2012znn}%
  \BibitemOpen
  \bibfield  {author} {\bibinfo {author} {\bibfnamefont {A.}~\bibnamefont
  {Bazavov}} \emph {et~al.} (\bibinfo {collaboration} {MILC}),\ }\href
  {\doibase 10.1103/PhysRevD.87.054505} {\bibfield  {journal} {\bibinfo
  {journal} {Phys. Rev. D}\ }\textbf {\bibinfo {volume} {87}},\ \bibinfo
  {pages} {054505} (\bibinfo {year} {2013})},\ \Eprint
  {http://arxiv.org/abs/1212.4768} {arXiv:1212.4768 [hep-lat]} \BibitemShut
  {NoStop}%
\bibitem [{\citenamefont {Hasenfratz}\ and\ \citenamefont
  {Knechtli}(2001)}]{Hasenfratz:2001hp}%
  \BibitemOpen
  \bibfield  {author} {\bibinfo {author} {\bibfnamefont {A.}~\bibnamefont
  {Hasenfratz}}\ and\ \bibinfo {author} {\bibfnamefont {F.}~\bibnamefont
  {Knechtli}},\ }\href {\doibase 10.1103/PhysRevD.64.034504} {\bibfield
  {journal} {\bibinfo  {journal} {Phys. Rev. D}\ }\textbf {\bibinfo {volume}
  {64}},\ \bibinfo {pages} {034504} (\bibinfo {year} {2001})},\ \Eprint
  {http://arxiv.org/abs/hep-lat/0103029} {arXiv:hep-lat/0103029} \BibitemShut
  {NoStop}%
\bibitem [{\citenamefont {Ji}\ \emph {et~al.}(2021{\natexlab{b}})\citenamefont
  {Ji}, \citenamefont {Liu}, \citenamefont {Sch\"afer}, \citenamefont {Wang},
  \citenamefont {Yang}, \citenamefont {Zhang},\ and\ \citenamefont
  {Zhao}}]{Ji:2020brr}%
  \BibitemOpen
  \bibfield  {author} {\bibinfo {author} {\bibfnamefont {X.}~\bibnamefont
  {Ji}}, \bibinfo {author} {\bibfnamefont {Y.}~\bibnamefont {Liu}}, \bibinfo
  {author} {\bibfnamefont {A.}~\bibnamefont {Sch\"afer}}, \bibinfo {author}
  {\bibfnamefont {W.}~\bibnamefont {Wang}}, \bibinfo {author} {\bibfnamefont
  {Y.-B.}\ \bibnamefont {Yang}}, \bibinfo {author} {\bibfnamefont {J.-H.}\
  \bibnamefont {Zhang}}, \ and\ \bibinfo {author} {\bibfnamefont
  {Y.}~\bibnamefont {Zhao}},\ }\href {\doibase 10.1016/j.nuclphysb.2021.115311}
  {\bibfield  {journal} {\bibinfo  {journal} {Nucl. Phys. B}\ }\textbf
  {\bibinfo {volume} {964}},\ \bibinfo {pages} {115311} (\bibinfo {year}
  {2021}{\natexlab{b}})},\ \Eprint {http://arxiv.org/abs/2008.03886}
  {arXiv:2008.03886 [hep-ph]} \BibitemShut {NoStop}%
\bibitem [{\citenamefont {Orginos}\ \emph {et~al.}(2017)\citenamefont
  {Orginos}, \citenamefont {Radyushkin}, \citenamefont {Karpie},\ and\
  \citenamefont {Zafeiropoulos}}]{Orginos:2017kos}%
  \BibitemOpen
  \bibfield  {author} {\bibinfo {author} {\bibfnamefont {K.}~\bibnamefont
  {Orginos}}, \bibinfo {author} {\bibfnamefont {A.}~\bibnamefont {Radyushkin}},
  \bibinfo {author} {\bibfnamefont {J.}~\bibnamefont {Karpie}}, \ and\ \bibinfo
  {author} {\bibfnamefont {S.}~\bibnamefont {Zafeiropoulos}},\ }\href {\doibase
  10.1103/PhysRevD.96.094503} {\bibfield  {journal} {\bibinfo  {journal} {Phys.
  Rev. D}\ }\textbf {\bibinfo {volume} {96}},\ \bibinfo {pages} {094503}
  (\bibinfo {year} {2017})},\ \Eprint {http://arxiv.org/abs/1706.05373}
  {arXiv:1706.05373 [hep-ph]} \BibitemShut {NoStop}%
\bibitem [{\citenamefont {Ji}\ \emph {et~al.}(2015)\citenamefont {Ji},
  \citenamefont {Sch\"afer}, \citenamefont {Xiong},\ and\ \citenamefont
  {Zhang}}]{Ji:2015qla}%
  \BibitemOpen
  \bibfield  {author} {\bibinfo {author} {\bibfnamefont {X.}~\bibnamefont
  {Ji}}, \bibinfo {author} {\bibfnamefont {A.}~\bibnamefont {Sch\"afer}},
  \bibinfo {author} {\bibfnamefont {X.}~\bibnamefont {Xiong}}, \ and\ \bibinfo
  {author} {\bibfnamefont {J.-H.}\ \bibnamefont {Zhang}},\ }\href {\doibase
  10.1103/PhysRevD.92.014039} {\bibfield  {journal} {\bibinfo  {journal} {Phys.
  Rev. D}\ }\textbf {\bibinfo {volume} {92}},\ \bibinfo {pages} {014039}
  (\bibinfo {year} {2015})},\ \Eprint {http://arxiv.org/abs/1506.00248}
  {arXiv:1506.00248 [hep-ph]} \BibitemShut {NoStop}%
\bibitem [{\citenamefont {Liu}\ \emph {et~al.}(2019)\citenamefont {Liu},
  \citenamefont {Wang}, \citenamefont {Xu}, \citenamefont {Zhang},
  \citenamefont {Zhao},\ and\ \citenamefont {Zhao}}]{Liu:2018tox}%
  \BibitemOpen
  \bibfield  {author} {\bibinfo {author} {\bibfnamefont {Y.-S.}\ \bibnamefont
  {Liu}}, \bibinfo {author} {\bibfnamefont {W.}~\bibnamefont {Wang}}, \bibinfo
  {author} {\bibfnamefont {J.}~\bibnamefont {Xu}}, \bibinfo {author}
  {\bibfnamefont {Q.-A.}\ \bibnamefont {Zhang}}, \bibinfo {author}
  {\bibfnamefont {S.}~\bibnamefont {Zhao}}, \ and\ \bibinfo {author}
  {\bibfnamefont {Y.}~\bibnamefont {Zhao}},\ }\href {\doibase
  10.1103/PhysRevD.99.094036} {\bibfield  {journal} {\bibinfo  {journal} {Phys.
  Rev. D}\ }\textbf {\bibinfo {volume} {99}},\ \bibinfo {pages} {094036}
  (\bibinfo {year} {2019})},\ \Eprint {http://arxiv.org/abs/1810.10879}
  {arXiv:1810.10879 [hep-ph]} \BibitemShut {NoStop}%
\bibitem [{\citenamefont {Ball}\ \emph {et~al.}(2007)\citenamefont {Ball},
  \citenamefont {Braun},\ and\ \citenamefont {Lenz}}]{Ball:2007zt}%
  \BibitemOpen
  \bibfield  {author} {\bibinfo {author} {\bibfnamefont {P.}~\bibnamefont
  {Ball}}, \bibinfo {author} {\bibfnamefont {V.~M.}\ \bibnamefont {Braun}}, \
  and\ \bibinfo {author} {\bibfnamefont {A.}~\bibnamefont {Lenz}},\ }\href
  {\doibase 10.1088/1126-6708/2007/08/090} {\bibfield  {journal} {\bibinfo
  {journal} {JHEP}\ }\textbf {\bibinfo {volume} {08}},\ \bibinfo {pages} {090}
  (\bibinfo {year} {2007})},\ \Eprint {http://arxiv.org/abs/0707.1201}
  {arXiv:0707.1201 [hep-ph]} \BibitemShut {NoStop}%
\bibitem [{\citenamefont {Roberts}\ \emph {et~al.}(2021)\citenamefont
  {Roberts}, \citenamefont {Richards}, \citenamefont {Horn},\ and\
  \citenamefont {Chang}}]{Roberts:2021nhw}%
  \BibitemOpen
  \bibfield  {author} {\bibinfo {author} {\bibfnamefont {C.~D.}\ \bibnamefont
  {Roberts}}, \bibinfo {author} {\bibfnamefont {D.~G.}\ \bibnamefont
  {Richards}}, \bibinfo {author} {\bibfnamefont {T.}~\bibnamefont {Horn}}, \
  and\ \bibinfo {author} {\bibfnamefont {L.}~\bibnamefont {Chang}},\ }\href
  {\doibase 10.1016/j.ppnp.2021.103883} {\bibfield  {journal} {\bibinfo
  {journal} {Prog. Part. Nucl. Phys.}\ }\textbf {\bibinfo {volume} {120}},\
  \bibinfo {pages} {103883} (\bibinfo {year} {2021})},\ \Eprint
  {http://arxiv.org/abs/2102.01765} {arXiv:2102.01765 [hep-ph]} \BibitemShut
  {NoStop}%
\bibitem [{\citenamefont {Edwards}\ and\ \citenamefont
  {Joo}(2005)}]{Edwards:2004sx}%
  \BibitemOpen
  \bibfield  {author} {\bibinfo {author} {\bibfnamefont {R.~G.}\ \bibnamefont
  {Edwards}}\ and\ \bibinfo {author} {\bibfnamefont {B.}~\bibnamefont {Joo}}
  (\bibinfo {collaboration} {SciDAC, LHPC, UKQCD}),\ }\bibfield  {booktitle}
  {\emph {\bibinfo {booktitle} {{Lattice field theory. Proceedings, 22nd
  International Symposium, Lattice 2004, Batavia, USA, June 21-26, 2004}}},\
  }\href {\doibase 10.1016/j.nuclphysbps.2004.11.254} {\bibfield  {journal}
  {\bibinfo  {journal} {Nucl. Phys. Proc. Suppl.}\ }\textbf {\bibinfo {volume}
  {140}},\ \bibinfo {pages} {832} (\bibinfo {year} {2005})},\ \bibinfo {note}
  {[,832(2004)]},\ \Eprint {http://arxiv.org/abs/hep-lat/0409003}
  {arXiv:hep-lat/0409003 [hep-lat]} \BibitemShut {NoStop}%
\bibitem [{\citenamefont {Clark}\ \emph {et~al.}(2010)\citenamefont {Clark},
  \citenamefont {Babich}, \citenamefont {Barros}, \citenamefont {Brower},\ and\
  \citenamefont {Rebbi}}]{Clark:2009wm}%
  \BibitemOpen
  \bibfield  {author} {\bibinfo {author} {\bibfnamefont {M.~A.}\ \bibnamefont
  {Clark}}, \bibinfo {author} {\bibfnamefont {R.}~\bibnamefont {Babich}},
  \bibinfo {author} {\bibfnamefont {K.}~\bibnamefont {Barros}}, \bibinfo
  {author} {\bibfnamefont {R.~C.}\ \bibnamefont {Brower}}, \ and\ \bibinfo
  {author} {\bibfnamefont {C.}~\bibnamefont {Rebbi}},\ }\href {\doibase
  10.1016/j.cpc.2010.05.002} {\bibfield  {journal} {\bibinfo  {journal}
  {Comput. Phys. Commun.}\ }\textbf {\bibinfo {volume} {181}},\ \bibinfo
  {pages} {1517} (\bibinfo {year} {2010})},\ \Eprint
  {http://arxiv.org/abs/0911.3191} {arXiv:0911.3191 [hep-lat]} \BibitemShut
  {NoStop}%
\bibitem [{\citenamefont {Babich}\ \emph {et~al.}(2011)\citenamefont {Babich},
  \citenamefont {Clark}, \citenamefont {Joo}, \citenamefont {Shi},
  \citenamefont {Brower},\ and\ \citenamefont {Gottlieb}}]{Babich:2011np}%
  \BibitemOpen
  \bibfield  {author} {\bibinfo {author} {\bibfnamefont {R.}~\bibnamefont
  {Babich}}, \bibinfo {author} {\bibfnamefont {M.~A.}\ \bibnamefont {Clark}},
  \bibinfo {author} {\bibfnamefont {B.}~\bibnamefont {Joo}}, \bibinfo {author}
  {\bibfnamefont {G.}~\bibnamefont {Shi}}, \bibinfo {author} {\bibfnamefont
  {R.~C.}\ \bibnamefont {Brower}}, \ and\ \bibinfo {author} {\bibfnamefont
  {S.}~\bibnamefont {Gottlieb}},\ }in\ \href {\doibase 10.1145/2063384.2063478}
  {\emph {\bibinfo {booktitle} {{SC11 International Conference for High
  Performance Computing, Networking, Storage and Analysis Seattle, Washington,
  November 12-18, 2011}}}}\ (\bibinfo {year} {2011})\ \Eprint
  {http://arxiv.org/abs/1109.2935} {arXiv:1109.2935 [hep-lat]} \BibitemShut
  {NoStop}%
\bibitem [{\citenamefont {Clark}\ \emph {et~al.}(2016)\citenamefont {Clark},
  \citenamefont {Jo}, \citenamefont {Strelchenko}, \citenamefont {Cheng},
  \citenamefont {Gambhir},\ and\ \citenamefont {Brower}}]{Clark:2016rdz}%
  \BibitemOpen
  \bibfield  {author} {\bibinfo {author} {\bibfnamefont {M.~A.}\ \bibnamefont
  {Clark}}, \bibinfo {author} {\bibfnamefont {B.}~\bibnamefont {Jo}}, \bibinfo
  {author} {\bibfnamefont {A.}~\bibnamefont {Strelchenko}}, \bibinfo {author}
  {\bibfnamefont {M.}~\bibnamefont {Cheng}}, \bibinfo {author} {\bibfnamefont
  {A.}~\bibnamefont {Gambhir}}, \ and\ \bibinfo {author} {\bibfnamefont
  {R.}~\bibnamefont {Brower}},\ }\href@noop {} {\  (\bibinfo {year} {2016})},\
  \Eprint {http://arxiv.org/abs/1612.07873} {arXiv:1612.07873 [hep-lat]}
  \BibitemShut {NoStop}%
\bibitem [{\citenamefont {Bi}\ \emph {et~al.}(2020)\citenamefont {Bi},
  \citenamefont {Xiao}, \citenamefont {Gong}, \citenamefont {Guo},
  \citenamefont {Sun}, \citenamefont {Xu},\ and\ \citenamefont
  {Yang}}]{Bi:2020wpt}%
  \BibitemOpen
  \bibfield  {author} {\bibinfo {author} {\bibfnamefont {Y.-J.}\ \bibnamefont
  {Bi}}, \bibinfo {author} {\bibfnamefont {Y.}~\bibnamefont {Xiao}}, \bibinfo
  {author} {\bibfnamefont {M.}~\bibnamefont {Gong}}, \bibinfo {author}
  {\bibfnamefont {W.-Y.}\ \bibnamefont {Guo}}, \bibinfo {author} {\bibfnamefont
  {P.}~\bibnamefont {Sun}}, \bibinfo {author} {\bibfnamefont {S.}~\bibnamefont
  {Xu}}, \ and\ \bibinfo {author} {\bibfnamefont {Y.-B.}\ \bibnamefont
  {Yang}},\ }\bibfield  {booktitle} {\emph {\bibinfo {booktitle} {{Proceedings,
  37th International Symposium on Lattice Field Theory (Lattice 2019): Wuhan,
  China, June 16-22 2019}}},\ }\href {\doibase 10.22323/1.363.0286} {\bibfield
  {journal} {\bibinfo  {journal} {PoS}\ }\textbf {\bibinfo {volume}
  {LATTICE2019}},\ \bibinfo {pages} {286} (\bibinfo {year} {2020})},\ \Eprint
  {http://arxiv.org/abs/2001.05706} {arXiv:2001.05706 [hep-lat]} \BibitemShut
  {NoStop}%
\bibitem [{\citenamefont {Hua}\ \emph {et~al.}(2021)\citenamefont {Hua},
  \citenamefont {Chu}, \citenamefont {Sun}, \citenamefont {Wang}, \citenamefont
  {Xu}, \citenamefont {Yang}, \citenamefont {Zhang},\ and\ \citenamefont
  {Zhang}}]{Hua:2020gnw}%
  \BibitemOpen
  \bibfield  {author} {\bibinfo {author} {\bibfnamefont {J.}~\bibnamefont
  {Hua}}, \bibinfo {author} {\bibfnamefont {M.-H.}\ \bibnamefont {Chu}},
  \bibinfo {author} {\bibfnamefont {P.}~\bibnamefont {Sun}}, \bibinfo {author}
  {\bibfnamefont {W.}~\bibnamefont {Wang}}, \bibinfo {author} {\bibfnamefont
  {J.}~\bibnamefont {Xu}}, \bibinfo {author} {\bibfnamefont {Y.-B.}\
  \bibnamefont {Yang}}, \bibinfo {author} {\bibfnamefont {J.-H.}\ \bibnamefont
  {Zhang}}, \ and\ \bibinfo {author} {\bibfnamefont {Q.-A.}\ \bibnamefont
  {Zhang}} (\bibinfo {collaboration} {Lattice Parton}),\ }\href {\doibase
  10.1103/PhysRevLett.127.062002} {\bibfield  {journal} {\bibinfo  {journal}
  {Phys. Rev. Lett.}\ }\textbf {\bibinfo {volume} {127}},\ \bibinfo {pages}
  {062002} (\bibinfo {year} {2021})},\ \Eprint
  {http://arxiv.org/abs/2011.09788} {arXiv:2011.09788 [hep-lat]} \BibitemShut
  {NoStop}%
\end{thebibliography}%

\end{document}